\documentclass[12pt]{amsart}
\usepackage[T1]{fontenc}
\usepackage{mathabx}
\usepackage{url}
\usepackage{graphicx}
\usepackage{hyperref}
\hypersetup{
    colorlinks=true,
    linkcolor=purple,
    citecolor=purple,
    urlcolor=blue,
}
\usepackage{booktabs}
\usepackage{longtable}
\usepackage{standalone}
\usepackage{pgfplots}
\pgfplotsset{compat=1.18}
\usepackage{tikz}
\usetikzlibrary{intersections, patterns}
\usepackage{amsmath, amsthm, amsfonts, amssymb}
\usepackage{amsaddr}
\usepackage{xspace}

\usepackage[margin=1.25in]{geometry}
\usepackage{mathtools}
\usepackage{cleveref}
\usepackage{subcaption}
\usepackage{cite}
\usepackage{enumitem}
\usepackage{microtype}

\usepackage[p,osf]{cochineal}
\usepackage[cochineal]{newtxmath}
\newtheorem{theorem}{Theorem}

\newtheorem{proposition}[theorem]{Proposition}

\newcommand{\lab}[1]{}

\newcommand{\toint}{sqrt(3)}

\newcommand{\varA}{\textsc{a}}
\newcommand{\varB}{\textsc{b}}
\newcommand{\varC}{\textsc{c}}
\newcommand{\varConv}{\textsc{conv}}

\newcommand{\imb}{\Delta}

\newcommand{\localizer}{\texttt{Localizer}}

\usepackage[noEnd,commentColor=blue,rightComments=false]{algpseudocodex}
\usepackage{algorithm}

\makeatletter
\def\orcidID#1{\href{http://orcid.org/#1}{\protect\raisebox{-1.25pt}{\protect\includegraphics{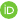}}}}
\makeatother

\usepackage{color}

\usepackage{xcolor}

\DeclareMathOperator{\conv}{conv}

\begin{document}
\title[Automated Symmetric Constructions in Discrete Geometry]{\large Automated Symmetric Constructions\\ in Discrete Geometry }
%
%
\author[B. Subercaseaux, E. Mackey, L. Qian, and M.J.H. Heule]{\small Bernardo Subercaseaux \orcidID{0000-0003-2295-1299},
Ethan Mackey \orcidID{0009-0005-9130-6797}, \\  Long Qian  \orcidID{0000-0003-1567-3948} \and
Marijn J. H. Heule \orcidID{0000-0002-5587-8801}}

%
%
\email{\{bsuberca,ethanmac,longq,mheule\}@andrew.cmu.edu}
\address{Carnegie Mellon University, Pittsburgh, PA 15213, USA}
%
%
\begin{abstract}
    We present a computational methodology for obtaining rotationally symmetric sets of points satisfying discrete geometric constraints, and demonstrate its applicability by discovering new solutions to some well-known problems in combinatorial geometry. Our approach takes the usage of SAT solvers in discrete geometry further by directly embedding rotational symmetry into the combinatorial encoding of geometric configurations.
    Then, to realize concrete point sets corresponding to abstract designs provided by a SAT solver, we introduce a novel local-search realizability solver, which shows excellent practical performance despite the intrinsic $\exists \mathbb{R}$-completeness of the problem. Leveraging this combined approach, we provide symmetric extremal solutions to the Erd\H{o}s-Szekeres problem, as well as a minimal odd-sized solution with 21 points for the everywhere-unbalanced-points problem, improving on the previously known 23-point configuration.
    The imposed symmetries yield more aesthetically appealing solutions, enhancing human interpretability, and simultaneously offer computational benefits by significantly reducing the number of variables required to encode discrete geometric problems.
\end{abstract}

\vspace*{-1cm}
\maketitle    


\section{Introduction}\label{sec:intro}

Symmetric solutions to combinatorial problems present several benefits: they tend to be easier to grasp and generalize~\cite{Giora2008-eq,Weyl2016-ed}, and can even be easier to compute since they have fewer degrees of freedom (i.e., optimization variables)~\cite{walshSymmetrySolutions2010,heuleSymmetrySolutions2010,heuleAvoidingTriplesArithmetic2017a}. In the words of Turing awardee Alan J. Perlis, \emph{``Symmetry is a complexity-reducing concept; seek it everywhere''}~\cite{Perlis1982}.

Despite theese benefits, it can be hard to prove in advance that a given problem will have symmetric solutions. This is especially the case in Ramsey theory, where the existence of objects avoiding certain patterns is often proven by (pseudo)random constructions, or asymmetric inductive arguments. For example, Schur number $k$, a classic Ramsey-theoretical problem, asks for the largest integer $n$ such that there exists a $k$-coloring of $\{1, \ldots, n\}$ with no monocromatic solution to $x+y=z$, and the following \emph{palindromic} $3$-coloring is optimal since $S(3) = 13$:
\[
    \textcolor{red}{1} \;
    \textcolor{green!60!black}{2} \; 
    \textcolor{green!60!black}{3} \; 
    \textcolor{red}{4} \; 
    \textcolor{blue}{5} \; 
    \textcolor{blue}{6} \; 
    \textcolor{green!60!black}{7} \; 
    \textcolor{blue}{8} \; 
    \textcolor{blue}{9} \; 
    \textcolor{red}{10} \; 
    \textcolor{green!60!black}{11} \; 
    \textcolor{green!60!black}{12} \; 
    \textcolor{red}{13}.
\]

Interestingly, for all known Schur numbers $S(k)$, there is an optimal coloring that is palindromic (symmetric w.r.t. $i \mapsto n\!-\!i\!+\!1$)~\cite{heuleSchurNumberFive2018}. However, this is not known to be true for all $k$.
A comparable phenomenon has been observed for van der Waerden numbers~\cite{heuleAvoidingTriplesArithmetic2017a}.
In this article we show that a similar situation seems to occur in certain discrete geometry problems, and that such symmetric solutions can be found \emph{automatically}.
We will consider two problems in discrete geometry described below.

\begin{figure}[t]
    \begin{subfigure}{0.32\textwidth}
        \centering
        \fbox{  \begin{tikzpicture}[dot/.style={inner sep=1pt, circle, minimum size=3mm, thick, scale=0.6}]
    \begin{axis}[
      unit vector ratio=1 1 1,
      xmin=0,xmax=17,
      ymin=-16.5, ymax=0,
      grid=both,
      grid style={line width=.1pt, draw=gray!10},
      major grid style={line width=.2pt,draw=gray!50},
      axis lines=none,
      minor tick num=1,
      axis line style={latex-latex},
      xticklabel=\empty,
      yticklabel=\empty ]
\node[dot, fill=blue!100!red] (1) at (axis cs:0.633333, -6.5) {\lab{1}};
\node[dot, fill=blue!100!red] (2) at (axis cs:8.46667, -0.7) {\lab{1}};
\node[dot, fill=blue!100!red] (3) at (axis cs:16.3333, -6.53333) {\lab{1}};
\node[dot, fill=blue!100!red] (4) at (axis cs:13.3667, -15.8333) {\lab{1}};
\node[dot, fill=blue!100!red] (5) at (axis cs:3.66667, -15.9333) {\lab{1}};
\node[dot, fill=blue!10!red] (6) at (axis cs:8.36367, -9.11667) {\lab{1}};
\node[dot, fill=blue!70!red] (7) at (axis cs:3.90067, -11.7247) {\lab{1}};
\node[dot, fill=blue!70!red] (8) at (axis cs:13.504, -12.614) {\lab{1}};
\node[dot, fill=blue!40!red] (9) at (axis cs:9.081, -4.67133) {\lab{1}};
\node[dot, fill=blue!70!red] (10) at (axis cs:3.064, -8.109) {\lab{1}};
\node[dot, fill=blue!70!red] (11) at (axis cs:13.783, -10.6577) {\lab{1}};
\node[dot, fill=blue!10!red] (12) at (axis cs:7.62133, -7.134) {\lab{1}};
\node[dot, fill=blue!70!red] (13) at (axis cs:8.54267, -3.121) {\lab{1}};
\node[dot, fill=blue!40!red] (14) at (axis cs:3.87233, -9.457) {\lab{1}};
\node[dot, fill=blue!10!red] (15) at (axis cs:8.179, -7.82267) {\lab{1}};
\node[dot, fill=blue!40!red] (16) at (axis cs:13.584, -11.2133) {\lab{1}};
    \end{axis}
  \end{tikzpicture}}
        \caption{Asymmetric construction of $16$ points in general position without a convex $6$-gon~\cite{wiki:Happy_ending_problem}.}
        \label{fig:wiki16}
    \end{subfigure}
    \hfill
    \begin{subfigure}{0.32\textwidth}
        \centering
        \fbox{ \begin{tikzpicture}[dot/.style={inner sep=1pt, circle, minimum size=1.5mm, thick, scale=1.1}, scale=0.3]
        \node[dot, fill=blue!10!red] at (4, 1.0) {};
        \node[dot, fill=blue!40!red] at (5, 0.953125) {};
        \node[dot, fill=blue!70!red] at (6, 0.9375) {};
        \node[dot, fill=blue!100!red] at (7, 0.953125) {};
        \node[dot, fill=blue!10!red] at (-4, -1.0) {};
        \node[dot, fill=blue!40!red] at (-5, -0.953125) {};
        \node[dot, fill=blue!70!red] at (-6, -0.9375) {};
        \node[dot, fill=blue!100!red] at (-7, -0.953125) {};
        \node[dot, fill=blue!10!red] at (-1.0, 4) {};
        \node[dot, fill=blue!40!red] at (-0.953125, 5) {};
        \node[dot, fill=blue!70!red] at (-0.9375, 6) {};
        \node[dot, fill=blue!100!red] at (-0.953125, 7) {};
        \node[dot, fill=blue!10!red] at (1.0, -4) {};
        \node[dot, fill=blue!40!red] at (0.953125, -5) {};
        \node[dot, fill=blue!70!red] at (0.9375, -6) {};
        \node[dot, fill=blue!100!red] at (0.953125, -7) {};

\end{tikzpicture} }
        \caption{4-fold symmetric construction that minimizes the number of convex pentagons~\cite{subercaseaux2024automatedmathematicaldiscoveryverification}.}
        \label{fig:pinwheel-construction}
    \end{subfigure}
    \hfill
    \begin{subfigure}{0.31\textwidth}
        \centering
        \fbox{
%
%
%
%
%

\begin{tikzpicture}[scale=0.6]
    \useasboundingbox (-2.8,-2.4) rectangle (2.6,3);



     \begin{scope}[shift={(-0.2, -0.15)}]
\tikzset{
    dot/.style={circle,inner sep=1pt,fill,label={\tiny #1},name=#1}}

    
\node[dot, fill=blue!10!red] (A) at (0,0) {};
\node[dot, fill=blue!10!red] (B) at (1,0) {};
\node[dot, fill=blue!10!red] (C) at (0.5,{0.5*sqrt(3)}) {};
\node[dot, fill=blue!70!red] (D) at (-1,0) {};
\node[dot, fill=blue!100!red] (E) at (-2,0) {};
\node[dot, fill=blue!70!red] (F) at (1.5,{-0.5*sqrt(3)}) {};
\node[dot, fill=blue!100!red] (G) at (2,{-1*sqrt(3)}) {};
\node[dot, fill=blue!70!red] (H) at (1,{1*sqrt(3)}) {};
\node[dot, fill=blue!100!red] (I) at (1.5,{1.5*sqrt(3)}) {};

\draw [name path=A--F, draw=none] (A) -- (F);
\draw [name path=B--H, draw=none] (B) -- (H);
\draw [name path=C--D, draw=none] (C) -- (D);

\coordinate (X) at (intersection of A--F and B--H);
\node[dot, fill=blue!40!red] (X1) at (X) {};

\coordinate (Y) at (intersection of A--F and C--D);
\node[dot, fill=blue!40!red] (Y1) at (Y) {};

\coordinate (Z) at (intersection of B--H and C--D);
\node[dot, fill=blue!40!red] (Z1) at (Z) {};

\draw (A) -- (B) -- (C) -- (A);
\draw (A) -- (D) -- (E);
\draw (B) -- (F) -- (G);
\draw (C) -- (H) -- (I);

\draw (X1) -- (B) -- (Z1) -- (H);
\draw (Y1) -- (A) -- (X1) -- (F);
\draw (Z1) -- (C) -- (Y1) -- (D);
\end{scope}
\end{tikzpicture}
}
        \caption{A 12-point 2-EU configuration with a 3-fold symmetry~\cite{ResearchProblems2002}. Lines connect collinear points.}
        \label{fig:alon-construction}
    \end{subfigure}
    \caption{Examples of geometric constructions with and without symmetry.}
\end{figure}
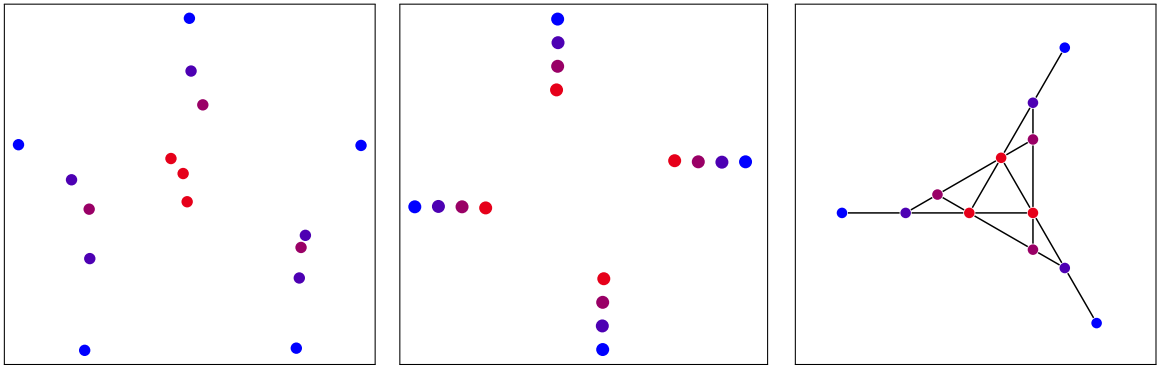

\paragraph{\textbf{Erd\H{o}s-Szekeres.}} For each integer $k \geq 3$, the problem is to find $g(k)$, the smallest integer such that any set of $g(k)$ points in the plane, without three on a common line, contains $k$ points in convex position. 
This long-standing problem dates back to the 1930s, with Klein's proof of $g(4) = 5$, popularly known as the \emph{happy-ending theorem}.  The only further values known are $g(5) = 9$ and $g(6) = 17$~\cite{szekeres_peters_2006}. Erd\H{o}s and Szekeres conjectured that $g(k) = 2^{k-2} + 1$ for every $k$, which matches the three known datapoints, and proved this to be a lower bound~\cite{morrisErdosSzekeresProblemPoints2000b}.
The Erd\H{o}s-Szekeres upper-bound construction is however asymmetric and hard to visualize even for small values of $k$. For example, an asymmetric construction of showing $g(6) > 16$, currently in the Wikipedia page of the Happy Ending problem~\cite{wiki:Happy_ending_problem}, is shown in~\Cref{fig:wiki16}. In contrast, our symmetric constructions are presented in~\Cref{sec:results} (\Cref{fig:16-6-symmetric}).
Interestingly, Subercaseaux et al.~\cite{subercaseaux2024automatedmathematicaldiscoveryverification} provided a symmetric construction for minimizing the number of convex pentagons amongst $n$ points, and both the work of Morris and Soltan~\cite{morrisErdosSzekeresProblemPoints2000b} as well as Scheucher~\cite{scheucherTwoDisjoint5holes2020a} exhibited symmetric (or almost symmetric) extremal constructions for variants of the Erd\H{o}s-Szekeres problem. In this work, we present a much more systematic approach to symmetry in discrete geometry.

\paragraph{\textbf{Everywhere-unbalanced-points.}} Given a set $S$ of $n$ points in the plane, we denote by $L(S)$ the set of lines that touch at least two points of $S$. The ``imbalance'' $\imb(\ell)$ of a line $\ell \in L(S)$ is the difference between the number of points that are on one side of $\ell$ and the number of points on the other side. We say that $S$ is $k$-everywhere-unbalanced ($k$-EU, for short) if $\imb(\ell) \geq k$ for all lines $\ell \in L(S)$. The main open problem is whether $k$-EU sets exist for all $k$. This question can be traced back to Kupitz, who asked whether a $2$-EU pointset exists~\cite{Kupitz}. Alon answered affirmatively (see~\Cref{fig:alon-construction}), showing that $2$-EU sets with $4s$ points exist for every odd value of $s \geq 3$, and his construction has an $s$-fold rotational symmetry~\cite{ResearchProblems2002}.
 Pinchasi proved that $k$-EU sets, if they exist, must have $\Omega(2^{2^{k}})$ points~\cite{pinchasiLinesManyPoints2003}, and recently Conlon and Lim showed an almost-matching upper bound when considering pseudolines instead of lines~\cite{conlon2025unbalancedconfigurations}. It is still open, however, whether a $k$-EU set exists for any $k \geq 3$.

\subsection{Our contributions and methodology}
We present a computational methodology for obtaining rotationally symmetric constructions for discrete geometric problems, which we apply to the aforementioned problems as well as other variants, e.g., $9$ points in general position without an empty convex $5$-gon. 
The methodology consists of two stages:
\begin{itemize}
    \setlength\itemsep{1em}
    \item[] \textbf{Symmetric combinatorial encoding:} We first encode the geometric properties of the problems as instances of boolean satisfiability (SAT), following a recent line of work. We impose rotational symmetry in the encoding itself, which was initially explored by one of the authors~\cite{EthanMSc}. To do this efficiently, we prove that the axioms of Knuth's CC systems~\cite{knuthAxiomsHulls1992} can be encoded with $O(n^4)$ many clauses without assuming a left-to-right ordering of the points as in previous work~\cite{heuleHappyEndingEmpty2024c}. For the everywhere-unbalanced-points problem, we use SAT for the first time and prove, for instance, that Alon's $12$-point construction uses the minimum number of points for a $2$-EU set. 
    \item[] \textbf{Realizability:} Going from the solutions to the SAT instances to actual pointsets turns out to be harder, both in theory and in practice. To tackle this we developed~\localizer, a local-search solver for the $\exists \mathbb{R}$-complete realizability problem, and through an experimental evaluation show it is highly performant compared to the general-purpose local-search formulation used in previous work~\cite{subercaseaux2024automatedmathematicaldiscoveryverification}. Our solver does not handle collinear points, and thus to realize solutions of the everywhere-unbalanced-points problem (which necessarily involves collinearity), we use a different ad-hoc approach.
\end{itemize}

We show symmetric solutions for the Erd\H{o}s-Szekeres problem on $16$ points without convex $6$-gons, and furthermore we enumerate and classify them, showing that $s$-fold symmetric solutions exist for $s = 4$ and $s = 5$, but not for $s = 3$. 
For the everywhere-unbalanced-points problem, we exhibit a $21$-point symmetric construction (\Cref{fig:unbalanced}), which we prove to be the minimal odd-sized solution. In this case, the symmetry crucially allows us to find a realization. 

\noindent
Our code is publicly available at \url{github.com/bsubercaseaux/automatic-symmetries}.

\section{Background}\label{sec:background}
    
While the domain of the aforementioned problems is continuous (i.e., $\mathbb{R}^2$), it is possible to reason about geometric properties like convexity, or line balances, purely in terms of combinatorial relationships between points.   
A widely successful abstraction in discrete geometry is that of \emph{triple orientations}~\cite{knuthAxiomsHulls1992} which consists of characterizing, for each ordered triples of points $(p, q, r)$, whether it defines a curve that turns clockwise, counterclockwise, or whether they are collinear. Concretely, given points \(p, q, r\), their triple orientation is defined as
\[
    \sigma(p, q, r) =  \text{sign} \det \begin{pmatrix} p_x & q_x & r_x \\ p_y & q_y & r_y \\ 1 & 1 & 1 \end{pmatrix} = \begin{cases} -1 & \text{if } p, q, r \text{ are oriented clockwise,} \\
    \;\; \, 0 & \text{if } p, q, r \text{ are collinear}, \\
       \;\; \, 1 & \text{if } p, q, r \text{ are oriented counterclockwise}.
    \end{cases}
\]

This abstraction has been successfully used to encode and solve several problems in discrete geometry: Peters and Szekeres used it to settle $g(6) = 17$~\cite{szekeres_peters_2006}, Heule and Scheucher for proving that $30$ points in general position must contain an empty convex hexagon~\cite{heuleHappyEndingEmpty2024c}, among others~\cite{scheucherTwoDisjoint5holes2020a,subercaseaux2024automatedmathematicaldiscoveryverification,scheucherSATAttackErdosSzekeres2023a}.
Let us clarify how these orientations are enough to express the constraints of both the Erd\H{o}s-Szekeres and the everywhere-unbalanced-points problems. We will start with some definitions. A set of points $S$ is in general position if no three points are collinear, that is, $\sigma(p, q, r) \neq 0$ for all $(p, q, r) \in \binom{S}{3}$. For a finite set of points $S \subset \mathbb{R}^2$, we denote by $\conv(S)$ the \emph{convex hull} of $S$, which is the smallest convex set containing $S$. 
Then, a set of points $S$ in general position is in \emph{convex position} if removing any point $a \in S$ would change its convex hull, i.e., $\conv\left(S \setminus  \{a\}\right) \neq \conv(S)$ for all $a \in S$. As a consequence of Carathéodory's theorem, a set of points $S$ in general position is in convex position if and only if every subset of $4$ points of $S$ is also in convex position. This implies that, by using the triple orientations to express whether sets of $4$ points are in convex position, we can express the presence of a convex $k$-gon (from now on, we will use $k$-gon to refer to a set of $k$ points in convex position). The precise formulation is in~\Cref{subsec:k-gon-constraints}, and we remark that a Lean formalization of these basic discrete geometry notions has been carried out by Subercaseaux et al.~\cite{subercaseauxFormalVerificationEmpty2024}.
For the everywhere-unbalanced-points problem, it will suffice to express the imbalance of a line $\ell$ between points $p$ and $q$ of a pointset $S$, as the absolute value of the difference
between
    $|\{r \in S : \sigma(p, q, r) = 1\}|$ and $|\{r \in S : \sigma(p, q, r) = -1\}|.$

\subsection{Geometric and combinatorial symmetries}

We consider two different forms of symmetry. For a pointset $P = \{p_1, \ldots, p_n\}$, a \emph{combinatorial symmetry} is a bijection $\pi: \{1, \ldots, n\} \to \{1, \ldots, n\}$ such that $\sigma(p_i, p_j, p_k) = \sigma(p_{\pi(i)}, p_{\pi(j)}, p_{\pi(k)})$ for all $(i, j, k) \in \binom{[n]}{3}$. On the other hand, a \emph{geometric symmetry} is a bijection $\rho: \mathbb{R}^2 \to \mathbb{R}^2$ such that $\rho(p_i) \in P$ for all $i \in \{1, \ldots, n\}$. We say that a geometric symmetry $\rho$ is \emph{orientation preserving} if $\sigma(\rho(p), \rho(q), \rho(r)) = \sigma(p, q, r)$ for every $p, q, r \in \mathbb{R}^2$. For example, any rotation of the plane is orientation preserving~\cite{subercaseauxFormalVerificationEmpty2024}, but it might not be a geometric symmetry of a pointset, as illustrated in~\Cref{fig:wiki16} taking $\pi/4$ as rotation angle.
In this work, we focus on $s$-fold rotational symmetries, which correspond to rotations of the plane by $\frac{2\pi}{s}$, for some integer $s$. More formally, let $\rho_\alpha : \mathbb{R}^2 \to \mathbb{R}^2$ be the function defined by $\rho_\alpha(x, y) = (x \cos(\alpha) - y \sin(\alpha), x \sin(\alpha) + y \cos(\alpha))$. Then, we say a set of points $S$ is \emph{$s$-fold symmetric} if $\rho_{2\pi/{s}}$ is a geometric symmetry of $S$ (i.e., $\rho_{2\pi/{s}}(S) = S$). 

\subsection{Orientation Variables}\label{sec:orientation_variables}

To obtain a propositional encoding for the aforementioned problems, we start by defining \emph{orientation variables} $\varA_{i, j, k}$, $\varB_{i, j, k}$, and $\varC_{i, j, k}$ for each triple of distinct indices $i, j, k \in \{1, \ldots, n\}$, where $n$ is the number of points in the desired pointset. 
$\varA_{i, j, k}$ will represent that $\sigma(p_i, p_j, p_k) = 1$, whereas $\varB_{i, j, k}$ that $\sigma(p_i, p_j, p_k) = -1$, and $\varC_{i, j, k}$ that $\sigma(p_i, p_j, p_k) = 0$. For the Erd\H{o}s-Szekeres problem, given that we are interested in pointsets in general position, we will only need the variables $\varA_{i, j, k}$, since its truth value is enough to identify the orientation of the triple $(p_i, p_j, p_k)$. For the everywhere-unbalanced-points problem, we use the three kinds of variables, and naturally enforce that exactly one of them is true for each triple of points.
From now on, we assume the number of points $n$ to be fixed.

\subsection{CC Systems and Axioms}\label{subsec:cc_systems}

The study of axioms for combinatorial representations of pointsets was initiated by Knuth~\cite{knuthAxiomsHulls1992}, who introduced the so-called \emph{CC systems}, as an abstraction for pointsets in general position. Knuth's axioms can be written as follows in our notation:
\begin{enumerate}[leftmargin=2cm]
    \item[\textbf{Axiom 1}] (Cyclic Symmetry). $\varA_{i, j, k} \rightarrow \varA_{j, k, i}$,
    \item[\textbf{Axiom 2}] (Antisymmetry). $\varA_{i, j, k} \rightarrow \overline{\varA_{i, k, j}}$,
    \item[\textbf{Axiom 3}] (Nondegeneracy). $\varA_{i, j, k} \lor \varA_{i, k, j}$,
    \item[\textbf{Axiom 4}] (Interiority). $\varA_{i, j, k} \lor \varA_{i, k, l} \lor \varA_{k, j, \ell} \lor \varA_{j, i, \ell}$,
    \item[\textbf{Axiom 5}] (Transitivity). $\varA_{\ell, i, m} \lor \varA_{\ell,j,m} \lor \varA_{\ell,k,m} \lor \varA_{\ell, j, i} \lor \varA_{\ell, k, j} \lor \varA_{\ell,i,k}$,
\end{enumerate}
where each axiom is quantified over all distinct indices $i, j, k, \ell, m \in \{1, \ldots, n\}$.
It turns out that Axioms (1-3)
can be encoded implicitly, by only using variables $\varA_{i, j, k}$ for indices $i < j < k$, replacing each occurrence of a variable whose indices are not ordered with the corresponding variable with ordered indices, and a potential negation. Namely, for a tuple $t$ of three indices not necessarily sorted, we can consider $t'$ as the sorted version of $t$. According to Axioms (1-3), we replace each occurrence of $\varA_t$ with $\varA_{t'}$ if $t$ has an even number of inversions (i.e., the number of swaps required to sort $t$), and with $\overline{\varA_{t'}}$ otherwise.
Despite these optimizations, the number of clauses required to encode the axioms is still roughly $5! \cdot \binom{n}{5} \approx n^5$, which amounts to over $24$ million clauses for $n = 32$.

\subsection{Signotope Axioms}\label{subsec:signotope_axiom}

A more efficient alternative in terms of encoding size is to use the so-called \emph{signotope axioms}~\cite{felsnerSweepsArrangementsSignotopes2001,scheucherTwoDisjoint5holes2020a,heuleHappyEndingEmpty2024c}, which assuming that points are sorted from left to right (i.e., $x_i < x_{i+1}$ for every $i$, where $x_i$ denotes the $x$-coordinate of point $p_i$), 
allows to express an equivalent set of axioms with only $4 \binom{n}{4}$ clauses. 
The signotope axioms can be written in clausal form as follows:
\begin{equation}
    (\overline{\varA_{i,j,k}} \lor \overline{\varA_{i,k,\ell}} \lor \varA_{i,j,\ell}) \land (\varA_{i,j,k} \lor \varA_{i,k,\ell} \lor \overline{\varA_{i,j,\ell}}),
\end{equation}
\begin{equation}
(\overline{\varA_{i,j,k}} \lor \overline{\varA_{j,k,\ell}} \lor \varA_{i,k,\ell}) \land (\varA_{i,j,k} \lor \varA_{j,k,\ell} \lor \overline{\varA_{i,k,\ell}}),
\end{equation}
where the quantification here is only over indices $1 \leq i < j < k < \ell \leq n$. The main issue with these signotope axioms, which we will address in~\Cref{sec:dynamic_point_ordering}, is that they assume a left-to-right ordering of the points, which can often be assumed without loss of generality by simply relabeling points from left to right (cf.~\cite{subercaseauxFormalVerificationEmpty2024}), but in our case is incompatible with the rotational symmetries we want to impose.
Note that the strictness of the ordering is not a restrictive condition since we can always rotate pointsets by an $\varepsilon$-angle while preserving all orientations to guarantee no points share the same $x$-coordinate. 

\subsection{Realizability Problem}

It is worth mentioning immediately that these combinatorial abstractions for pointsets are an \emph{under-approximation} of the geometric properties of points in $\mathbb{R}^2$, meaning that every set of points satisfies the axioms, but there are assignments of the orientation variables that satisfy the axioms and yet do not correspond to any planar pointset.
Therefore, if after adding problem-specific constraints (e.g., convexity, imbalance, etc.) we obtain a satisfiable formula, we still need to check whether the solution to the orientation variables can be realized in $\mathbb{R}^2$, the so-called \emph{point realizability problem} for which we present a local-search solver in~\Cref{sec:realizability}. On the other hand, if no assignment of the orientation variables satisfies the constraints of a problem in conjunction with the axioms, then we can safely conclude that no pointset exists satisfying the desired properties. This idea has been formalized in Lean by Subercaseaux et al.~\cite{subercaseauxFormalVerificationEmpty2024}.

\section{Symmetry Constraints}
\label{sec:symmetry}

In this section, we present the symmetry constraints that we will use to enforce $s$-fold rotational symmetry in our combinatorial encodings. 
In general, a combinatorial symmetry $\pi: \{1, \ldots, n\} \to \{1, \ldots, n\}$ can be enforced by adding clauses for the constraints
\(
    \varA_{i, j, k} \leftrightarrow \varA_{\pi(i), \pi(j), \pi(k)}, 
\)
and similarly for the $\varB_{i,j,k}$ and $\varC_{i,j,k}$ variables when dealing with collinear points. As we will see next, however, it is possible to enforce the symmetry directly without those clauses by unifying equivalent variables as we did for Axioms (1-2) of Knuth's CC systems.
For example, if we want to enforce a $4$-fold rotational symmetry over $16$ points, we can assume that the permutation $\pi$ can be factored as
\begin{equation}\label{eq:perm_factoring}
    \pi = (1,2,3,4)\,(5,6,7,8)\,(9,10,11,12)\, (13,14,15,16),
\end{equation}
which can be succinctly coded by defining $\pi(i) = \lfloor (i-1)/4 \rfloor \cdot 4 + (i+1)\! \! \mod 4$.
Then, if we consider the triple of indices $(1, 6, 8)$, we have the equivalences:
\[
    \varA_{1, 6, 8} \leftrightarrow \varA_{2, 7, 5} \leftrightarrow \varA_{3, 8, 6} \leftrightarrow \varA_{4, 5, 7}, 
\]
and by the equivalences of~\Cref{subsec:cc_systems}, this is the same as 
\[
    \varA_{1, 6, 8} \leftrightarrow \overline{\varA_{2, 5, 7}} \leftrightarrow \overline{\varA_{3, 6, 8}} \leftrightarrow \varA_{4, 5, 7}.
\]
We can thus treat these literals as an equivalence class, and for each such equivalence class we can choose a representative (e.g., the lexicographically smallest one, $\varA_{1, 6, 8}$), and then replace all occurrences of the other literals in its class with the representative or its negation. 

\subsection{Filtering Isomorphic Constraints}
Not only can we reduce the number of variables by the enforced symmetries, but also the number of constraints. For example, a constraint stating that indices $\{1, 3, 6, 8, 10, 13\}$ do not form a $6$-gon, already implies that the indices $\{2, 4, 5, 7, 11, 14\}$ do not form a $6$-gon when enforcing the symmetry of~\Cref{eq:perm_factoring}. Therefore, the second constraint is redundant and can be removed. In general, for a constraint involving a tuple of indices $t = (i_1, \ldots, i_k)$, we consider the \emph{orbit} of $t$ under the symmetry $\pi$, which is the set of all tuples of indices that can be obtained from $t$ by applying $\pi$ any number of times. We can then remove all but one constraint for each orbit, which we implement by only adding constraints that are lexicographically smallest in their orbits.

\subsection{Symmetry Breaking}

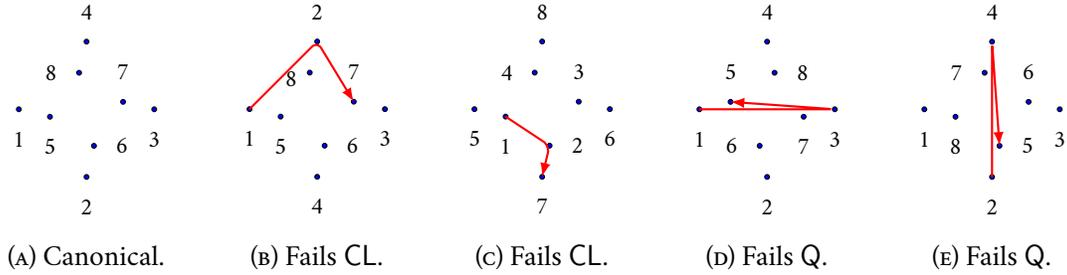
\begin{figure}[t]
    \begin{subfigure}{0.2\textwidth}
\centering
\begin{tikzpicture}[scale=0.18]
 \draw[circle, fill=blue] (-5, 0) circle (5pt); 
 \draw[circle, fill=blue] (5, 0) circle (5pt); 
 \draw[circle, fill=blue] (0, -5) circle (5pt); 
 \draw[circle, fill=blue] (0, 5) circle (5pt); 
 \node[label=below:{\scriptsize $1$}] at (-5, 0) {};
 \node[label=below:{\scriptsize $3$}] at (5, 0) {};
 \node[label=below:{\scriptsize $2$}] at (0, -5) {};
 \node[label=above:{\scriptsize $4$}] at (0, 5) {};

 \draw[circle, fill=blue] (-2.70, -0.55)  circle (5pt);
 \draw[circle, fill=blue] (0.55, -2.70)  circle (5pt);
 \draw[circle, fill=blue] (2.70, 0.55) circle (5pt);
 \draw[circle, fill=blue] (-0.55, 2.70)  circle (5pt);

 \node[label=below:{\scriptsize $5$}] at (-2.70, -0.55) {};
 \node[label=right:{\scriptsize $6$}] at (0.55, -2.70) {};
 \node[label=above:{\scriptsize $7$}] at (2.70, 0.55) {};
 \node[label=left:{\scriptsize $8$}] at (-0.55, 2.70) {};
 \end{tikzpicture}
 \caption{Canonical.}\label{fig:sym-canonical}
\end{subfigure}
\begin{subfigure}{0.19\textwidth}
\centering
\begin{tikzpicture}[scale=0.18]
 \draw[circle, fill=blue] (-5, 0) circle (5pt); 
 \draw[circle, fill=blue] (5, 0) circle (5pt); 
 \draw[circle, fill=blue] (0, -5) circle (5pt); 
 \draw[circle, fill=blue] (0, 5) circle (5pt); 
 \node[label=below:{\scriptsize $1$}] at (-5, 0) {};
 \node[label=below:{\scriptsize $3$}] at (5, 0) {};
 \node[label=below:{\scriptsize $4$}] at (0, -5) {};
 \node[label=above:{\scriptsize $2$}] at (0, 5) {};

 \draw[circle, fill=blue] (-2.70, -0.55)  circle (5pt);
 \draw[circle, fill=blue] (0.55, -2.70)  circle (5pt);
 \draw[circle, fill=blue] (2.70, 0.55) circle (5pt);
 \draw[circle, fill=blue] (-0.55, 2.70)  circle (5pt);

 \node[label=below:{\scriptsize $5$}] at (-2.70, -0.55) {};
 \node[label=right:{\scriptsize $6$}] at (0.55, -2.70) {};
 \node[label=above:{\scriptsize $7$}] at (2.70, 0.55) {};

 \draw[thick, -latex, red, rounded corners=1mm] (-5, 0) to (0, 5) to (2.70, 0.55);

  \node[label=left:{\scriptsize $8$}] at (0.15, 2.25) {};
 \end{tikzpicture}
 \caption{Fails \textsf{CL}.}\label{fig:fail1}
\end{subfigure}
\begin{subfigure}{0.19\textwidth}
\centering
\begin{tikzpicture}[scale=0.18]
 \draw[circle, fill=blue] (-5, 0) circle (5pt); 
 \draw[circle, fill=blue] (5, 0) circle (5pt); 
 \draw[circle, fill=blue] (0, -5) circle (5pt); 
 \draw[circle, fill=blue] (0, 5) circle (5pt); 
 \node[label=below:{\scriptsize $5$}] at (-5, 0) {};
 \node[label=below:{\scriptsize $6$}] at (5, 0) {};
 \node[label=below:{\scriptsize $7$}] at (0, -5) {};
 \node[label=above:{\scriptsize $8$}] at (0, 5) {};

 \draw[circle, fill=blue] (-2.70, -0.55)  circle (5pt);
 \draw[circle, fill=blue] (0.55, -2.70)  circle (5pt);
 \draw[circle, fill=blue] (2.70, 0.55) circle (5pt);
 \draw[circle, fill=blue] (-0.55, 2.70)  circle (5pt);

 \node[label=below:{\scriptsize $1$}] at (-2.70, -0.55) {};
 \node[label=right:{\scriptsize $2$}] at (0.55, -2.70) {};
 \node[label=above:{\scriptsize $3$}] at (2.70, 0.55) {};
 \node[label=left:{\scriptsize $4$}] at (-0.55, 2.70) {};

 \draw[thick, -latex, red, rounded corners=1mm] (-2.70, -0.55) to (0.55, -2.70) to (0, -5);
 \end{tikzpicture}
  \caption{Fails \textsf{CL}.}\label{fig:fail2}
\end{subfigure}
\begin{subfigure}{0.19\textwidth}
\centering
\begin{tikzpicture}[scale=0.18]
 \draw[circle, fill=blue] (-5, 0) circle (5pt); 
 \draw[circle, fill=blue] (5, 0) circle (5pt); 
 \draw[circle, fill=blue] (0, -5) circle (5pt); 
 \draw[circle, fill=blue] (0, 5) circle (5pt); 
 \node[label=below:{\scriptsize $1$}] at (-5, 0) {};
 \node[label=below:{\scriptsize $3$}] at (5, 0) {};
 \node[label=below:{\scriptsize $2$}] at (0, -5) {};
 \node[label=above:{\scriptsize $4$}] at (0, 5) {};


 \draw[circle, fill=blue] (-2.70, 0.55)  circle (5pt);
 \draw[circle, fill=blue] (-0.55, -2.70)  circle (5pt);
 \draw[circle, fill=blue] (2.70, -0.55) circle (5pt);
 \draw[circle, fill=blue] (0.55, 2.70)  circle (5pt);

 \node[label=above:{\scriptsize $5$}] at (-2.70, 0.55) {};
 \node[label=left:{\scriptsize $6$}] at (-0.55, -2.70) {};
 \node[label=below:{\scriptsize $7$}] at (2.70, -0.55) {};
 \node[label=right:{\scriptsize $8$}] at (0.55, 2.70) {};

 \draw[thick, -latex, red, rounded corners=1mm] (-5, 0) to (5, 0) to (-2.70, 0.55);
 \end{tikzpicture}
   \caption{Fails \textsf{Q}.}\label{fig:fail3}
\end{subfigure}
\begin{subfigure}{0.19\textwidth}
\centering
\begin{tikzpicture}[scale=0.18]
 \draw[circle, fill=blue] (-5, 0) circle (5pt); 
 \draw[circle, fill=blue] (5, 0) circle (5pt); 
 \draw[circle, fill=blue] (0, -5) circle (5pt); 
 \draw[circle, fill=blue] (0, 5) circle (5pt); 
 \node[label=below:{\scriptsize $1$}] at (-5, 0) {};
 \node[label=below:{\scriptsize $3$}] at (5, 0) {};
 \node[label=below:{\scriptsize $2$}] at (0, -5) {};
 \node[label=above:{\scriptsize $4$}] at (0, 5) {};

 \draw[circle, fill=blue] (-2.70, -0.55)  circle (5pt);
 \draw[circle, fill=blue] (0.55, -2.70)  circle (5pt);
 \draw[circle, fill=blue] (2.70, 0.55) circle (5pt);
 \draw[circle, fill=blue] (-0.55, 2.70)  circle (5pt);

 \node[label=below:{\scriptsize $8$}] at (-2.70, -0.55) {};
 \node[label=right:{\scriptsize $5$}] at (0.55, -2.70) {};
 \node[label=above:{\scriptsize $6$}] at (2.70, 0.55) {};
 \node[label=left:{\scriptsize $7$}] at (-0.55, 2.70) {};

 \draw[thick, -latex, red, rounded corners=1mm] (0, -5.0) to (0, 5) to (0.55, -2.70);
 \end{tikzpicture}
      \caption{Fails \textsf{Q}.}\label{fig:fail4}
\end{subfigure}
    \caption{Symmetry breaking predicates for pointsets with a $4$-fold symmetry. Orientations in red are failing to satisfy the symmetry-breaking predicates.}\label{fig:symmetry_breaking}
\end{figure}

To limit the generation of isomorphic solutions, we add symmetry-breaking predicates, which depend on the parameters of the problem at hand. Let us consider, for example, the Erd\H{o}s-Szekeres problem for $16$ points without $6$-gons and a $4$-fold symmetry.
The only possibility for the convex layers of these $16$ points is that we have $4$ layers with $4$ points each. Then, as any rotational symmetry must map each point $p$ to a point $q$ in the same convex layer as $p$ (potentially $p = q$), we can assume without loss of generality that the $4$-fold symmetry is precisely the one in~\Cref{eq:perm_factoring}.
Moreover, we can assume without loss of generality that the outermost convex layer is $\{1, 2, 3, 4\}$, that the next one is $\{5, 6, 7, 8\}$, and so on. To enforce that points $5, 6, 7, 8$ are inside the convex hull of points $1, 2, 3, 4$, and that $1\to2\to3\to4$ is a counterclockwise sequence, we can add \emph{convex layer unit clauses} (\textsf{CL}-clauses) of the form $\varA_{i, 1 + i \!\! \mod 4, j}$ for $i \in \{1,2,3,4\}$ and $j \in \{5, 6, 7, 8\}$. \Cref{fig:symmetry_breaking} illustrates how these \textsf{CL}-clauses enforece a \emph{canonical} representative from a set of isomorphic solutions. We can then add analogous \textsf{CL}-clauses to enforce that points $9, 10, 11, 12$ are inside the convex hull of points $5, 6, 7, 8$, and so on.
Furthermore, we can assume without loss of generality that all points whose index is $1 \! \! \mod 5$ are in the bottom-left quadrant. To see this, note that otherwise we could relabel the points assigning index $4i+1$ to whichever point from the $i$-th outermost convex layer is in the bottom left quadrant, noting that the $4$-fold symmetry enforces that at least one point per layer lies in that quadrant. 
Concretely, we add \emph{quadrant clauses} (\textsf{Q}-clauses) of the form $\overline{\varA_{1, 3, i}}$ and $\varA_{2, 4, i}$ for every $i > 1$ that is $1$ modulo $4$. Note that, assuming without loss of generality that point $1$ gets coordinates $(-C, 0)$, and that points $2, 3, 4$ get the coordinates implied by the orbit $1 \to 2 \to 3 \to 4$, then these clauses directly correspond to the points $4i + 1$ for $i \geq 1$ being in the bottom-left quadrant. See~\Cref{fig:fail3,fig:fail4}.


\section{Encodings}\label{sec:encodings}

In this section, we detail the propositional encodings for the Erd\H{o}s-Szekeres and everywhere-unbalanced-points problems. 

\subsection{Dynamic Point Ordering Axioms}\label{sec:dynamic_point_ordering}

While the signotope axioms are more efficient than the CC system axioms (an $n^5 \to n^4$ advantage), they assume a left-to-right ordering of the points, which is not compatible with the rotational symmetries we want to impose --- e.g., in the example symmetry of~\Cref{eq:perm_factoring} we can infer what the convex layers of the desired pointset are, but we cannot a priori say whether point $5$ will be to the left of point $12$ or to its right. 
A discovery of independent interest is that the left-to-right ordering can be replaced by any linear ordering $\prec$ of the point indices, and only apply the signotope axioms for tuples of indices respecting a constraint similar to $i \prec j \prec k \prec \ell$. That is, as opposed to a predefined ordering, we introduce variables $\prec_{i, j}$ for every pair of distinct indices, which the SAT solver will assign dynamically, and enforce axioms accordingly. Naturally, we need to add constraints stating that these variables induce a strict linear ordering:
\begin{enumerate}[leftmargin=4cm]
    \item[\textbf{Totality.}] $({\prec_{i, j}} \lor {\prec_{j, i}})$, for all $1 \leq i \neq j \leq n$.
    \item[\textbf{Asymmetry.}] $({\prec_{i, j}} \leftrightarrow \overline{{\prec_{j, i}}})$, for all $1 \leq i \neq j \leq n$.
    \item[\textbf{Transitivity.}] $({\prec_{i, j}} \land {\prec_{j, k}} \rightarrow {\prec_{i, k}})$, for all $1 \leq i \neq j \neq k \leq n$.
\end{enumerate}

These linear-ordering axioms incur in $\Theta(n^3)$ clauses, and thus are not a bottleneck. Then, it turns out that we can replace the signotope axioms for pointsets in general position with the following:
\begin{equation}\label{eq:dynamic_axiom1}
    \left(\prec_{i, j} \land \prec_{i, k} \land \prec_{i, \ell}\right) \rightarrow \left(\varA_{i, j, k} \lor \overline{\varA_{i, j, \ell}} \lor \varA_{i, k, \ell}\right), \quad \forall i \neq j \neq k \neq \ell,
\end{equation}
\begin{equation}\label{eq:dynamic_axiom2}
     \left(\prec_{i, k} \land \prec_{j, k} \land \prec_{k, \ell}\right) \rightarrow \left(\varA_{i, j, k} \lor \overline{\varA_{i, k, \ell}} \lor \varA_{j, k, \ell}\right), \quad \forall i \neq j \neq k \neq \ell.
\end{equation}
Moreover,~\Cref{eq:dynamic_axiom1} is only needed when $\max(j, k) < \ell$, which further reduces the number of clauses. In total, since the condition $\max(j, k) < \ell$ holds in exactly a third of the cases, these dynamic-ordering axioms incur in $\frac{4}{3} \cdot 4! \cdot \binom {n}{4} \approx \frac{4}{3} n^4$ clauses.
For $n=32$, this is around $1.3$ million clauses, a significant improvement over the $24$ million clauses of the CC-system axioms (cf.~\Cref{subsec:cc_systems}). We now summarize correctness with the following two propositions.
\begin{proposition}
    \label{prop:dynamic_axiom1}
    For every set of $n$ points $S = \{p_1, \ldots, p_n\}$ in general position with distinct $x$-coordinates, the assignment of the $\varA_{i, j, k}$ and ${\prec_{i, j}}$ variables $\tau$ defined as:
    \[
        \tau(\varA_{i, j, k}) = \begin{cases}
            \text{true} & \text{if } \sigma(p_i, p_j, p_k) = 1 \\
            \text{false} & \text{otherwise},
        \end{cases} \quad \tau(\prec_{i, j}) = \begin{cases}
            \text{true} & \text{if } x_i < x_j \\
            \text{false} & \text{otherwise},
        \end{cases}
    \]
    satisfies the dynamic-ordering axioms from~\Cref{eq:dynamic_axiom1,eq:dynamic_axiom2}.
\end{proposition}
\begin{proposition}\label{prop:dynamic_axiom2}
    For every assignment $\tau$ of the $\varA_{i, j, k}$ variables, with $i, j, k \in \{1, \ldots, n\}$, if there is an assignment $\theta$ to the ${\prec_{i, j}}$ variables ($1 \leq i \neq j \leq n$) such that $\tau \cup \theta$ satisfies the dynamic-ordering axioms from~\Cref{eq:dynamic_axiom1,eq:dynamic_axiom2}, and CC-Axioms (1-3) (see \Cref{subsec:cc_systems}), then 
    $\tau$ satisfies the CC-Axioms (4-5).
\end{proposition}
\Cref{prop:dynamic_axiom1} is stating that the dynamic-ordering axioms are respected by actual pointsets, and~\Cref{prop:dynamic_axiom2} is Intuitively stating that these axioms are no more permisive than the CC axioms. In other words, an empty set of axioms would trivially satisfy~\Cref{prop:dynamic_axiom1} but not~\Cref{prop:dynamic_axiom2}, and on the other hand, an inconsistent set of axioms would trivially satisfy~\Cref{prop:dynamic_axiom2} but not~\Cref{prop:dynamic_axiom1}.
Both proofs are included in~\Cref{sec:dynamic_axiom1_proof}. The first proof is algebraic, and similar to the proof of the signotope axioms in~\cite{subercaseauxFormalVerificationEmpty2024}, whereas the second proof is computational, since it reduces to the case $n = 5$.

\subsection{Axioms for Collinear Point Sets}\label{subsec:collinear_axioms}
In addition to the axioms in~\Cref{sec:dynamic_point_ordering}, additional care is needed to handle potentially collinear points. 
Intuitively, such axioms capture the property that for any collection of 4 points $p_i, p_j, p_k, p_\ell$ where the first 3 are collinear (i.e. $\varC_{i, j, k}$ is true), then the orientation of any triple that includes the point $p_\ell$ uniquely determines the orientation of all triples. 
As the points are not necessarily ordered, we leverage the dynamic point orderings and case on all possibilities. 
That is, we add the following clauses for every set of four distinct indices $i, j, k, \ell \in \{1, \ldots, n\}$:

\begin{enumerate}
    \item $(\varC_{i, j, k}) \rightarrow (\varC_{\mathbf{t_1}} \leftrightarrow \varC_{\mathbf{t_2}})$, for distinct triples $\mathbf{t_1}, \mathbf{t_2} \neq \{i, j, k\}$, 
    \item $({\prec_{i, j}} \land {\prec_{j, k}} \land \varC_{i, j, k}) \rightarrow (\varA_{\mathbf{t_1}} \leftrightarrow \varA_{\mathbf{t_2}})$, for distinct triples $\mathbf{t_1}, \mathbf{t_2} \neq \{i, j, k\}$, 
    \item $({\prec_{i, k}} \land {\prec_{k, j}} \land \varC_{i, j, k}) \rightarrow \left((\varA_{i, j, \ell} \leftrightarrow \varA_{i, k,  \ell}) \land  (\varA_{i, j,  \ell} \leftrightarrow \varB_{j, k,  \ell})\right)$,
    \item $({\prec_{k, i}} \land {\prec_{i, j}} \land \varC_{i, j, k}) \rightarrow \left((\varA_{i, j, \ell} \leftrightarrow \varB_{i, k,  \ell}) \land (\varA_{i, j,  \ell} \leftrightarrow \varB_{j, k,  \ell})\right)$.
\end{enumerate}
Moreover, for Axioms (2-4), we also add an analogous version with each $\prec$ variable negated, to consider the opposite ordering.
Axiom (1) enforces the transitivity of collinearity: if there exist two distinct triples of $\{p_i, p_j, p_k, p_l\}$ that are collinear, then all points are collinear. Axioms (2-4) handle the case where $\{p_i, p_j, p_k\}$ are collinear yet $p_\ell$ does not lie on the same line. In Axiom (2), either $i \prec j \prec k$ or $i \succ j \succ k$ hold, thus the points are ordered monotonically. As such, their relative orientations with respect to $p_\ell$ are necessarily equivalent, hence the bi-implication clauses. In contrast, consider the scenario where $i \prec k \prec j$, which falls under Axiom (3). In this case, the orientations $\sigma(p_i, p_j, p_\ell) = \sigma(p_i, p_k, p_\ell)$ are still equivalent, as the lines $p_i \to p_j, p_i \to p_k$ have the same direction, resulting in the clauses $\varA_{i, j, \ell} \leftrightarrow \varA_{i, k, \ell}$. On the contrary, the lines $p_i \to p_j, p_j \to p_k$ are now in opposing directions, thus the orientations $\sigma(p_i, p_j, p_\ell) = -\sigma(p_j, p_k, p_\ell)$ are opposites, yielding the clauses $\varA_{i, j, \ell} \leftrightarrow \varB_{j, k, \ell}$. Axiom (4) is identical and considers the remaining cases. 
Note that it suffices to add these clauses only for $i < j < k$, and $\ell \in [n] \setminus \{i, j, k\}$. Therefore, the total number of clauses is $28(n-3) \cdot \binom{n}{3} = 112 \binom{n}{4}$. 

\subsection{Constraints for $k$-Gons}\label{subsec:k-gon-constraints}

Similarly to the work of Scheucher~\cite{scheucherTwoDisjoint5holes2020a,scheucherSATAttackErdosSzekeres2023a}, we create auxiliary variables $\varConv_{i, j, k, \ell}$ for each set of four indices $i, j, k, \ell \in \{1, \ldots, n\}$ (note that these are unordered), representing whether the points $p_i, p_j, p_k, p_\ell$ are in convex position. Our encoding is slightly different from Scheucher's in that we define these variables only in terms of the base orientation variables:
\[
    \varConv_{i, j, k, \ell} \leftrightarrow ((\varA_{i, j, k} \leftrightarrow \varA_{i, k, \ell}) \leftrightarrow (\varA_{i, k, \ell} \leftrightarrow \varA_{j, k, \ell})),
\]
which we can express in $12 \binom{n}{4}$ clauses using Tseitin variables. The correctness of this encoding can be seen by a tedious case analysis of the $16$ possible orientations for the four points $p_i, p_j, p_k, p_\ell$. 
Then, as described in~\Cref{sec:background}, a set of points $S$ is in convex position if and only if every subset of $4$ points of $S$ is also in convex position. Thus, to express the absence of a convex $k$-gon, we enforce for every set $X \subseteq \{1, \ldots, n\}$ with $|X| = k$, the clause
\(
    \bigvee_{i, j, k, \ell \in X} \overline{\varConv_{i, j, k, \ell}}.
\)

\subsection{Constraints for Imbalance}
Recall that for a set of $n$ points $S$ in the plane, $L(S)$ denotes the set of all lines touching at least two points of $S$. For a pair of points $p_i, p_j \in S$, let $l_{i, j} \in L(S)$ be the unique line passing through $p_i$ and $p_j$. The set of points above/below $l_{i, j}$ are defined as
\(l^{+}_{i, j} = \{p_k \in S \mid \sigma(p_i, p_j, p_k) > 0\}\) and \(l^{-}_{i, j} = \{p_k \in S \mid \sigma(p_i, p_j, p_k) < 0\}\), respectively.
and the imbalance of $p_i, p_j$, denoted $\imb(i, j)$, is defined as
\(\imb(i, j) = ||l^{+}_{i, j}| - |l^{-}_{i, j}||.\)
Finally, the imbalance of $S$ is the minimum of all such imbalances \(\imb(S) = \min_{i, j \in {[n]\choose{2}}} \imb(i, j).\)
Therefore, to encode $\imb(S) \geq c$, it suffices to encode $\imb(i, j) \geq c$ for all pairs $i, j$.  Thus, the problem reduces to encoding that
\[
    \sum_{k \not \in \{i, j\}} \varA_{i, j, k} \geq \sum_{k \not \in \{i, j\}} \varB_{i, j, k} + c  \qquad \text{or} \qquad \sum_{k \not \in \{i, j\}} \varA_{i, j, k} \leq \sum_{k \not \in \{i, j\}} \varB_{i, j, k} - c.
\]
To achieve this, we use the standard Sinz encoding \cite{sinz2005encoding} to define for every pair $i, j$ the \emph{counting variables} $s_{m}$ and $t_{m}$, for each $0 \leq m \leq n-2$, which represent $\sum_{k \not \in \{i, j\}} \varA_{i, j, k} = m$ and $\sum_{k \not \in \{i, j\}} \varB_{i, j, k} = m$ respectively.   
Then, to encode that $\imb(i, j) \geq c$ it suffices to add clauses of the form
\(
    \overline{s}_{x} \lor \overline{t}_{y} 
\)
for each $x, y \in \{0, \ldots, n-2\}$ such that $|x - y| < c$.



\section{Realizability}\label{sec:realizability}
In the realizability problem, we are given an orientation assignment $\tau : \binom{n}{3} \to \{-1, 0, 1\}$, and the goal is to construct a set of points $S = \{p_1, \ldots, p_n\} \subset \mathbb{R}^2$ such that $\tau(i, j, k) = \sigma(p_i, p_j, p_k)$ for all $i < j < k$. 
This problem is known to be $\exists \mathbb{R}$-complete~\cite{tothHandbookDiscreteComputational2017}, and thus at least NP-hard (recall that $\textrm{NP} \subseteq \exists \mathbb{R} \subseteq \textrm{PSPACE}$~\cite{schaeferComplexityGeometricTopological2010}).
We first describe our efficient general-purpose realizability solver for pointsets in general position, and then the ad-hoc method we followed to realize pointsets with collinearities for the everywhere-unbalanced-points problem.

\subsection{\localizer: a Realizability Solver for Points in General Position}

Our solver (called \localizer) is written in C and it implements a local-search algorithm that starts with a random initialization of point coordinates $(x_1, y_1), \ldots, (x_n, y_n)$, and iteratively moves the points trying to minimize the number of orientation constraints that are not satisfied by their current positions.
The solver is multithreaded, and uses a parallelism model in which a global table of top-$K$ solutions is maintained, and threads independently select a random solution from this table (with probability proportional to the solution quality) and attempt to improve upon it.
If a thread manages to find a strictly better solution, then it updates the global table in case the found solution is better-or-equal than some solution on the top-$K$ table.
Within a thread, the algorithm uses a form of~\emph{simulated annealing}, where in each iteration a point is selected, with probabilities proportional to the number of unsatisfied constraints they participate in, and then the point is moved to a random position within a ball of radius $r$ centered at its previous position. The radius $r$ is exponentially decreasing. The pseudocode of the algorithm is presented in~\Cref{sec:local_realizer_pseudocode}.
Moreover,~\localizer~can receive the description of a rotational symmetry and find solutions that respect it.

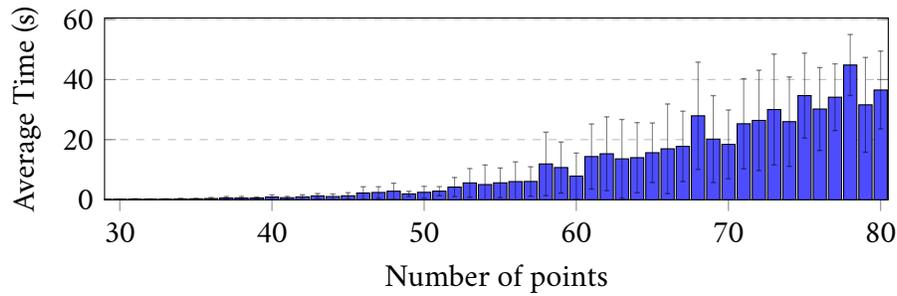
\begin{figure}[ht]
\centering
\begin{tikzpicture}[scale=1.0]
\begin{axis}[
    width=12cm,
    height=4cm,
    xlabel={Number of points},
    ylabel={Average Time (s)},
    xtick={30,40,...,80},
    xtick pos=bottom,
    xmax=80.5,
    xmin=30-1,
    ymin=0,
    ybar,
    bar width=5pt,
    ymajorgrids=true,
    grid style=dashed,
    error bars/y dir=both,
    error bars/y explicit,
]
\addplot[fill=blue!70, error bars/.cd, y dir=both, y explicit, error mark options={line width=0.5pt,mark size=1pt,rotate=90},  error bar style={line width=0.4pt, opacity=0.5}] coordinates {
    (20, 0.04) +- (0.07470398144709693, 0.07470398144709693)
    (21, 0.03) +- (0.02589062106668814, 0.02589062106668814)
    (22, 0.03) +- (0.006829735752300322, 0.006829735752300322)
    (23, 0.03) +- (0.005494803502807945, 0.005494803502807945)
    (24, 0.04) +- (0.015712649695723172, 0.015712649695723172)
    (25, 0.05) +- (0.018983035523003355, 0.018983035523003355)
    (26, 0.07) +- (0.03400857831691273, 0.03400857831691273)
    (27, 0.08) +- (0.0328522587658331, 0.0328522587658331)
    (28, 0.1) +- (0.05292293413817984, 0.05292293413817984)
    (29, 0.11) +- (0.05700722272381552, 0.05700722272381552)
    (30, 0.14) +- (0.06169199544152957, 0.06169199544152957)
    (31, 0.21) +- (0.15343405783816844, 0.15343405783816844)
    (32, 0.2) +- (0.08232272498352605, 0.08232272498352605)
    (33, 0.23) +- (0.11070585553590229, 0.11070585553590229)
    (34, 0.29) +- (0.16024683333646814, 0.16024683333646814)
    (35, 0.32) +- (0.12303644635423829, 0.12303644635423829)
    (36, 0.4) +- (0.3510132671038287, 0.3510132671038287)
    (37, 0.6) +- (0.4681968130351904, 0.4681968130351904)
    (38, 0.58) +- (0.5193756069707196, 0.5193756069707196)
    (39, 0.6) +- (0.2851073456765809, 0.2851073456765809)
    (40, 0.9) +- (0.705021643135006, 0.705021643135006)
    (41, 0.68) +- (0.43354895106198865, 0.43354895106198865)
    (42, 0.96) +- (0.6913083249529894, 0.6913083249529894)
    (43, 1.29) +- (0.7804776553717079, 0.7804776553717079)
    (44, 1.06) +- (0.8361000742549083, 0.8361000742549083)
    (45, 1.32) +- (1.0270026825435823, 1.0270026825435823)
    (46, 2.25) +- (2.0677137973633877, 2.0677137973633877)
    (47, 2.42) +- (1.905845071133149, 1.905845071133149)
    (48, 2.84) +- (2.6646995782226854, 2.6646995782226854)
    (49, 1.94) +- (0.8722192989493506, 0.8722192989493506)
    (50, 2.47) +- (1.9934960645382673, 1.9934960645382673)
    (51, 2.86) +- (1.4981928659207775, 1.4981928659207775)
    (52, 4.22) +- (3.1833678884868455, 3.1833678884868455)
    (53, 5.61) +- (4.772781330757802, 4.772781330757802)
    (54, 5.05) +- (6.507991432569825, 6.507991432569825)
    (55, 5.63) +- (4.936892458749421, 4.936892458749421)
    (56, 6.04) +- (6.560830851668417, 6.560830851668417)
    (57, 6.08) +- (4.888387158013487, 4.888387158013487)
    (58, 11.92) +- (10.55978733065466, 10.55978733065466)
    (59, 10.71) +- (8.478996672655635, 8.478996672655635)
    (60, 7.87) +- (7.666534339753452, 7.666534339753452)
    (61, 14.37) +- (10.826044357222166, 10.826044357222166)
    (62, 15.3) +- (12.281303734199769, 12.281303734199769)
    (63, 13.61) +- (13.10196213579365, 13.10196213579365)
    (64, 14.0) +- (11.673802529801991, 11.673802529801991)
    (65, 15.65) +- (9.900789713912904, 9.900789713912904)
    (66, 16.95) +- (14.918175655903871, 14.918175655903871)
    (67, 17.75) +- (11.725962898370913, 11.725962898370913)
    (68, 27.93) +- (17.86223434788224, 17.86223434788224)
    (69, 20.18) +- (14.459093753111524, 14.459093753111524)
    (70, 18.43) +- (11.449956600437218, 11.449956600437218)
    (71, 25.28) +- (15.018273161729383, 15.018273161729383)
    (72, 26.4) +- (16.690993887205448, 16.690993887205448)
    (73, 30.02) +- (18.44925826409947, 18.44925826409947)
    (74, 25.98) +- (14.874085985055997, 14.874085985055997)
    (75, 34.66) +- (14.138834811828227, 14.138834811828227)
    (76, 30.19) +- (13.794045780787096, 13.794045780787096)
    (77, 34.12) +- (11.102923454789742, 11.102923454789742)
    (78, 44.83) +- (10.15855199023246, 10.15855199023246)
    (79, 31.58) +- (15.765947597181778, 15.765947597181778)
    (80, 36.51) +- (12.955067308051445, 12.955067308051445)
};
\end{axis}
\end{tikzpicture}
\caption{Experimental evaluation of the~\localizer~realizability solver, over orientations obtained from random realizable sets of points (independently uniform in $[0, 1]^2$). Previous approaches took up to 100s for 16 points~\cite{subercaseaux2024automatedmathematicaldiscoveryverification}.}
\label{fig:realizer_bench}
\end{figure}

\paragraph{Evaluation.} We evaluate our algorithm on two families of realizable instances. On the one hand, we use the database of realizable order types for $n \leq 10$ points built by Aichholzer et al.~\cite{aichholzerEnumeratingOrderTypes2002}, where our algorithm solves every instance in less than $50$ milliseconds. On the other hand, we evaluate on randomly generated instances of $n$ points in general position, where we sample the points uniformly at random from the unit square (we tested other distributions and obtained similar results).  We conducted the experiments on a personal computer (MacBook Pro M1 2020, 16GB RAM, 8 CPU cores) using $8$ threads, and present the results in~\Cref{fig:realizer_bench}. In terms of performance,~\localizer~is roughly $6$ orders of magnitude faster than the local-search procedure described in our previous work~\cite{subercaseaux2024automatedmathematicaldiscoveryverification}, which could not solve any instance with $n = 30$ points in less than a minute.
We validated each solution through an independent Python program, which converts the floating-point coordinates to exact rational coordinates, and checks that all orientations are satisfied exactly.

\subsection{Realizing Collinear Configurations for Everywhere-Unbalanced}
As collinearity is an \emph{exact} condition where arbitrarily small perturbations will result in non-collinear points, it is not clear how purely numerical methods over the variables $\{x_1, y_1, \ldots, x_n, y_n\}$ can be used to satisfy the desired orientation constraints exactly. 
In fact, instead of trying to directly realize the orientation assignment $\tau$ obtained through SAT solving, we take $\tau$ as partial information to construct a set of points with the same imbalance, but not necessarily respecting the same orientations.
In particular, we start by extracting from $\tau$ a family of abstract lines $\mathcal{L}$ where each $L \in \mathcal{L}$ is a maximal set of indices (so $L \subseteq \{1, \ldots, n\}$) such that $\tau(i, j, k) = 0$ for all $i, j, k \in \binom{L}{3}$. 
Then, we note that if there are lines $L_1, L_2 \in \mathcal{L}$ such that $L_1 \cap L_2 = \{i\}$ for some $i \in \{1, \ldots, n\}$, then the point $p_i$ is fully determined by the remaining points (as two lines in Euclidean space have a unique intersection point), and we will say it is \emph{dependent}. 
After this, we aim to realize the remaining independent variables by maximizing the resulting unbalancedness using a numerical global optimization algorithm.
That is, we write the imbalance of each pair of points as a function of the independent variables, and then maximize the minimum imbalance over all pairs of points, using the $\texttt{differential\_evolution}$ algorithm implemented in SciPy~\cite{scipy}.





\section{Results}\label{sec:results}

For all our experiments, we used a version of the solver CaDiCaL\footnote{Available at \url{https://github.com:jreeves3/allsat-cadical}.}~\cite{CaDiCaL}
that is extended to support efficient enumeration of all solutions. 

\begin{figure}[t]
    \begin{subfigure}{0.49\textwidth}
        \centering
        \fbox{  \begin{tikzpicture}[dot/.style={inner sep=1pt, circle, minimum size=3mm, thick, scale=0.6}]
    \begin{axis}[
      unit vector ratio=1 1 1,
      xmin=-33.0,xmax=33.0,
      ymin=-33.0,ymax=33.0,
      grid=both,
      grid style={line width=.1pt, draw=gray!10},
      major grid style={line width=.2pt,draw=gray!50},
      axis lines=none,
      minor tick num=1,
      axis line style={latex-latex},
      xticklabel=\empty,
      yticklabel=\empty ]
\node[dot, fill=blue!100!red] (1) at (axis cs:-30.0, 0.0) {\lab{1}};
\node[dot, fill=blue!100!red] (2) at (axis cs:0.0, -30.0) {\lab{2}};
\node[dot, fill=blue!100!red] (3) at (axis cs:30.0, 0.0) {\lab{3}};
\node[dot, fill=blue!100!red] (4) at (axis cs:0.0, 30.0) {\lab{4}};
\node[dot, fill=blue!75!red] (5) at (axis cs:-20.0, -3.5) {\lab{5}};
\node[dot, fill=blue!75!red] (6) at (axis cs:3.5, -20.0) {\lab{6}};
\node[dot, fill=blue!75!red] (7) at (axis cs:20.0, 3.5) {\lab{7}};
\node[dot, fill=blue!75!red] (8) at (axis cs:-3.5, 20.0) {\lab{8}};
\node[dot, fill=blue!50!red] (9) at (axis cs:-13., -6.) {\lab{9}};
\node[dot, fill=blue!50!red] (10) at (axis cs:6., -13.) {\lab{10}};
\node[dot, fill=blue!50!red] (11) at (axis cs:13., 6.) {\lab{11}};
\node[dot, fill=blue!50!red] (12) at (axis cs:-6., 13.) {\lab{12}};
\node[dot, fill=blue!25!red] (13) at (axis cs:-1.9, -1.2) {\lab{13}};
\node[dot, fill=blue!25!red] (14) at (axis cs:1.2, -1.9) {\lab{14}};
\node[dot, fill=blue!25!red] (15) at (axis cs:1.9, 1.2) {\lab{15}};
\node[dot, fill=blue!25!red] (16) at (axis cs:-1.2, 1.9) {\lab{16}};

 \draw (1)--(5)--(9) -- (13);
 \draw (2)--(6)--(10) -- (14);
 \draw (3)--(7)--(11) -- (15);
 \draw (4)--(8)--(12) -- (16);
    \end{axis}
  \end{tikzpicture}}
        \caption{4-fold symmetric.}
        \label{fig:16-4fold}
    \end{subfigure}
    \hfill
    \begin{subfigure}{0.49\textwidth}
        \centering
        \fbox{\documentclass{standalone}
\usepackage{tikz}
\usepackage{pgfplots}

\newif\iflabel


\newcommand{\lab}[1]{}

\begin{document}
  \begin{tikzpicture}[dot/.style={inner sep=1pt, circle, minimum size=3mm, thick, scale=0.7}]
    \begin{axis}[
      unit vector ratio=1 1 1,
      xmin=-21,xmax=21,
      ymin=-19,ymax=22,
      grid=both,
      grid style={line width=.1pt, draw=gray!10},
      major grid style={line width=.2pt,draw=gray!50},
      axis lines=none,
      minor tick num=1,
      axis line style={latex-latex},
      xticklabel=\empty,
      yticklabel=\empty ]
\node[dot, fill=blue!100!red] (1) at (axis cs:-19.876200, 6.159400) {\lab{1}};
\node[dot, fill=blue!100!red] (2) at (axis cs:-12.000000, -17.000000) {\lab{1}};
\node[dot, fill=blue!100!red] (3) at (axis cs:12.459800, -16.666000) {\lab{1}};
\node[dot, fill=blue!100!red] (4) at (axis cs:19.700600, 6.699900) {\lab{1}};
\node[dot, fill=blue!100!red] (5) at (axis cs:-0.284100, 20.806700) {\lab{1}};
\node[dot, fill=blue!70!red] (6) at (axis cs:-15.000000, 2.000000) {\lab{1}};
\node[dot, fill=blue!70!red] (7) at (axis cs:-6.537400, -13.647800) {\lab{1}};
\node[dot, fill=blue!70!red] (8) at (axis cs:10.959700, -10.434800) {\lab{1}};
\node[dot, fill=blue!70!red] (9) at (axis cs:13.310800, 7.198700) {\lab{1}};
\node[dot, fill=blue!70!red] (10) at (axis cs:-2.733100, 14.883900) {\lab{1}};
\node[dot, fill=blue!40!red] (11) at (axis cs:-13.000000, 0.000000) {\lab{1}};
\node[dot, fill=blue!40!red] (12) at (axis cs:-4.017200, -12.363700) {\lab{1}};
\node[dot, fill=blue!40!red] (13) at (axis cs:10.517200, -7.641200) {\lab{1}};
\node[dot, fill=blue!40!red] (14) at (axis cs:10.517200, 7.641200) {\lab{1}};
\node[dot, fill=blue!40!red] (15) at (axis cs:-4.017200, 12.363700) {\lab{1}};
\node[dot, fill=blue!10!red] (16) at (axis cs:0.000000, 0.000000) {\lab{1}};

   \draw (1)--(6)--(11)--(16);
   \draw (2)--(7)--(12)--(16);
   \draw (3)--(8)--(13)--(16);
   \draw (4)--(9)--(14)--(16);
   \draw (5)--(10)--(15)--(16);
    \end{axis}
  \end{tikzpicture}
\end{document}}
        \caption{5-fold symmetric.}
        \label{fig:16-5fold}
    \end{subfigure}
    \caption{Constructions of $16$ points in general position without a convex $6$-gon.}
    \label{fig:16-6-symmetric}
\end{figure}
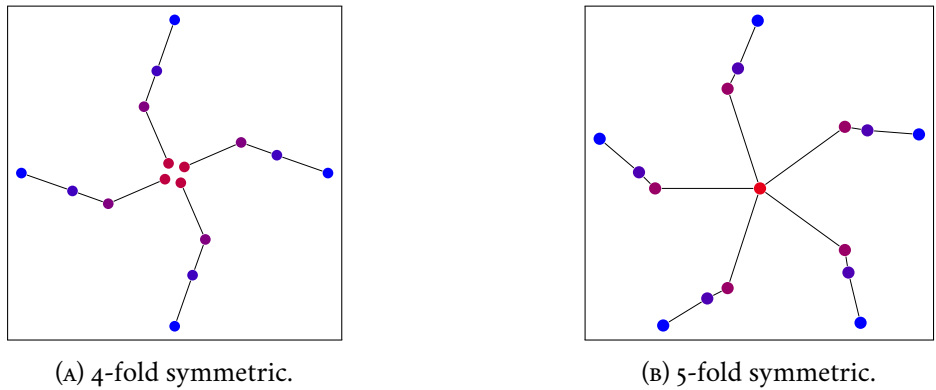

\subsection{Avoiding Hexagons}


We searched for symmetric configurations with 16 points that avoid hexagons (the maximum number where this can occur).
We constructed the formulas for $s \in \{3,4,5\}$. The formula enforcing a $3$-fold symmetry is unsatisfiable, while
the formulas with a $4$-fold and $5$-fold symmetry are satisfiable. After symmetry breaking, they have 66 and 
948 satisfying assignments, respectively. Computing all satisfying assignments can be achieved in a couple of
seconds using a single core on a personal computer (MackBook Pro M1 2020). 

Not all these 16-point satisfying assignments are realizable. In fact, most of them are (likely) unrealizable\footnote{The realizer tool can only
determine realizability, not unrealizability. However, for several of the configurations for which the tool was unable to find a realization, we were able to find a subset 10-point configuration that is known to be unrealizable.}. Out of the
66 solutions of the $4$-fold symmetry, 18 are realizable, while out of the 932 solutions of the $5$-fold symmetry
92 are realizable. Figure~\ref{fig:16-6-symmetric} shows a $4$-fold and a $5$-fold realization. 

\begin{figure}[t]
\centering
\begin{tikzpicture}
\begin{axis}[%
            xlabel=4-gons,
            ylabel=5-gons,
            axis y line=left,
           axis x line = bottom,
           width=3cm,
           xtick = {924},
           height=9cm,
           ymin=200,
            ymax=325,
                enlargelimits=false,
                        xmin=924,
            xmax=924,
            scatter/classes={%
                a={mark=*, fill=blue, draw=blue, fill opacity=0.5, draw opacity=0},
                b={mark=*, fill=red, draw=red, fill opacity=0.8, draw opacity=0}}]
\addplot[scatter, only marks, scatter src=explicit symbolic]%
table[meta=label] {
x y label
924 232 a
924 232 a
924 244 a
924 256 a
924 260 a
924 280 a
924 280 a
924 280 a
924 288 a
924 288 a
924 292 a
924 296 a
924 296 a
924 296 a
924 304 a
924 304 a
924 304 a
924 304 a
924 304 a
924 304 a
924 304 a
924 308 a
924 320 a
924 320 a
924 208 b
924 208 b
924 228 b
924 244 b
924 248 b
924 272 b
924 272 b
924 280 b
924 296 b
    };
\end{axis}
\end{tikzpicture}
\begin{tikzpicture}
\begin{axis}[%
            legend style={at={(0.45,0.9)}},
            xlabel=4-gons,
           width=12cm,
           height=9cm,
           axis y line=left,
           axis x line = bottom,
            ymin=200,
            xmin=780,
            xmax=1210,
            scatter/classes={%
                a={mark=*, fill=blue, draw=blue, fill opacity=0.5, draw opacity=0},
                b={mark=*, fill=red, draw=red, fill opacity=0.8, draw opacity=0}}]
\addplot[scatter, only marks, scatter src=explicit symbolic]%
table[meta=label] {
x y label
1000 508 a
1000 523 a
1000 523 a
1000 523 a
1000 533 a
1000 533 a
1000 543 a
1000 548 a
1000 563 a
1000 573 a
1000 573 a
1000 573 a
1000 583 a
1000 593 a
1000 603 a
1000 603 a
1000 608 a
1000 628 a
1005 528 a
1005 548 a
1005 548 a
1005 558 a
1005 558 a
1005 568 a
1005 588 a
1005 598 a
1005 618 a
1005 628 a
1010 553 a
1010 563 a
1010 573 a
1010 583 a
1010 613 a
1015 548 a
1015 553 a
1015 568 a
1015 573 a
1015 588 a
1020 568 a
1020 573 a
1020 573 a
1020 583 a
1020 583 a
1020 588 a
1020 603 a
1020 648 a
1020 668 a
1025 668 a
1025 698 a
1025 718 a
1035 593 a
1035 593 a
1035 598 a
1035 598 a
1035 608 a
1035 613 a
1035 613 a
1035 633 a
1035 633 a
1035 633 a
1035 658 a
1035 678 a
1040 613 a
1040 613 a
1040 613 a
1040 623 a
1040 628 a
1040 628 a
1040 643 a
1050 633 a
1050 633 a
1050 653 a
1055 628 a
1055 638 a
1055 648 a
1055 658 a
1055 658 a
1055 668 a
1055 668 a
1055 673 a
1055 693 a
1060 643 a
1060 653 a
1060 663 a
1060 663 a
1060 683 a
1060 693 a
1065 643 a
1065 648 a
1065 678 a
1075 713 a
1075 733 a
1080 668 a
1080 668 a
1080 673 a
1080 673 a
1080 683 a
1080 688 a
1080 688 a
1080 708 a
1080 708 a
1080 708 a
1080 733 a
1080 733 a
1080 753 a
1080 763 a
1080 783 a
1085 688 a
1085 698 a
1085 703 a
1085 703 a
1085 718 a
1095 708 a
1095 708 a
1095 728 a
1100 728 a
1100 733 a
1100 743 a
1100 743 a
1100 748 a
1100 768 a
1105 738 a
1105 758 a
1105 768 a
1110 753 a
1115 758 a
1115 758 a
1115 818 a
1120 788 a
1120 788 a
1120 788 a
1120 808 a
1125 808 a
1125 838 a
1125 858 a
1130 793 a
1135 828 a
1135 828 a
1140 823 a
1140 843 a
1140 853 a
1145 838 a
1155 873 a
1155 888 a
1155 893 a
1160 893 a
1160 923 a
1160 923 a
1160 923 a
1160 943 a
1180 968 a
1180 988 a
1185 1018 a
1185 1038 a
1185 988 a
800 263 a
800 263 a
800 293 a
800 293 a
800 303 a
800 308 a
800 308 a
800 328 a
800 343 a
800 348 a
825 263 a
825 263 a
825 298 a
825 318 a
825 323 a
825 328 a
825 328 a
825 338 a
825 358 a
825 363 a
835 298 a
835 318 a
835 328 a
840 263 a
840 263 a
840 303 a
840 313 a
840 348 a
840 348 a
840 348 a
840 353 a
840 368 a
840 368 a
840 383 a
840 388 a
840 388 a
840 388 a
845 263 a
845 308 a
845 358 a
845 358 a
845 358 a
845 368 a
845 368 a
845 368 a
845 378 a
845 383 a
845 403 a
860 303 a
860 323 a
860 328 a
860 333 a
860 343 a
860 348 a
860 353 a
860 373 a
860 403 a
860 413 a
870 383 a
870 383 a
875 308 a
875 318 a
875 338 a
875 358 a
875 368 a
875 368 a
875 378 a
875 378 a
875 388 a
875 408 a
875 418 a
875 418 a
875 428 a
875 428 a
880 313 a
880 348 a
880 373 a
880 383 a
880 388 a
880 388 a
880 393 a
880 403 a
880 403 a
880 413 a
880 443 a
885 343 a
885 358 a
885 378 a
885 398 a
885 408 a
885 418 a
885 428 a
890 403 a
890 413 a
890 433 a
900 353 a
900 373 a
900 373 a
900 383 a
900 388 a
900 403 a
900 408 a
900 408 a
900 408 a
900 423 a
900 423 a
900 433 a
900 448 a
900 448 a
900 448 a
900 468 a
905 363 a
905 388 a
905 388 a
905 403 a
905 408 a
905 408 a
905 408 a
905 408 a
905 418 a
905 418 a
905 423 a
905 438 a
905 438 a
905 443 a
905 448 a
905 448 a
915 398 a
915 408 a
915 428 a
920 403 a
920 413 a
920 428 a
920 428 a
920 428 a
920 428 a
920 433 a
920 433 a
920 433 a
920 443 a
920 443 a
920 448 a
920 448 a
920 448 a
920 463 a
920 468 a
920 468 a
920 468 a
920 468 a
925 418 a
925 438 a
925 448 a
925 448 a
925 448 a
925 448 a
925 448 a
925 458 a
925 463 a
925 483 a
930 423 a
930 433 a
930 443 a
935 438 a
935 453 a
935 453 a
940 453 a
940 463 a
940 473 a
940 473 a
940 473 a
940 483 a
940 493 a
945 443 a
945 458 a
945 458 a
945 458 a
945 468 a
945 468 a
945 468 a
945 478 a
945 478 a
945 483 a
945 498 a
945 498 a
945 508 a
945 508 a
945 508 a
945 518 a
950 463 a
950 483 a
950 483 a
950 493 a
950 493 a
950 503 a
955 488 a
955 498 a
955 508 a
955 518 a
960 453 a
960 468 a
960 473 a
960 473 a
960 473 a
960 483 a
960 483 a
960 483 a
960 488 a
960 503 a
960 503 a
960 508 a
960 508 a
960 513 a
960 523 a
965 503 a
965 508 a
965 508 a
965 518 a
965 518 a
965 523 a
965 538 a
975 468 a
975 493 a
975 493 a
975 493 a
975 498 a
975 498 a
975 513 a
975 513 a
975 528 a
975 533 a
975 533 a
975 533 a
975 533 a
980 483 a
980 513 a
980 513 a
980 513 a
980 523 a
980 528 a
980 528 a
980 528 a
980 533 a
980 533 a
980 543 a
980 548 a
980 548 a
980 548 a
980 568 a
980 568 a
980 568 a
980 593 a
980 613 a
985 488 a
985 498 a
985 508 a
985 548 a
985 558 a
985 563 a
985 563 a
985 578 a
995 538 a
995 538 a
995 568 a
995 568 a
995 588 a
1000 583 b
1005 538 b
1010 543 b
1010 603 b
1015 558 b
1020 573 b
1020 573 b
1025 658 b
1035 593 b
1040 603 b
1040 603 b
1040 613 b
1055 618 b
1055 648 b
1060 633 b
1065 648 b
1065 668 b
1080 668 b
1080 723 b
1085 678 b
1085 688 b
1100 723 b
1105 728 b
1110 743 b
1125 798 b
1140 813 b
1145 828 b
1160 883 b
1185 978 b
870 373 b
885 388 b
905 418 b
920 428 b
925 438 b
940 463 b
950 473 b
955 478 b
960 493 b
965 508 b
965 508 b
975 493 b
980 503 b
980 528 b
985 538 b
985 548 b
995 528 b
    };
\legend{unrealizable,realizable}
\end{axis}
\end{tikzpicture}
\caption{The number of 4-gons and 5-gons in realizable and (likely) unrealizable configurations with a 4-fold symmetry (left) and a 5-fold symmetry (right).}
\label{fig:scatter}
\end{figure}
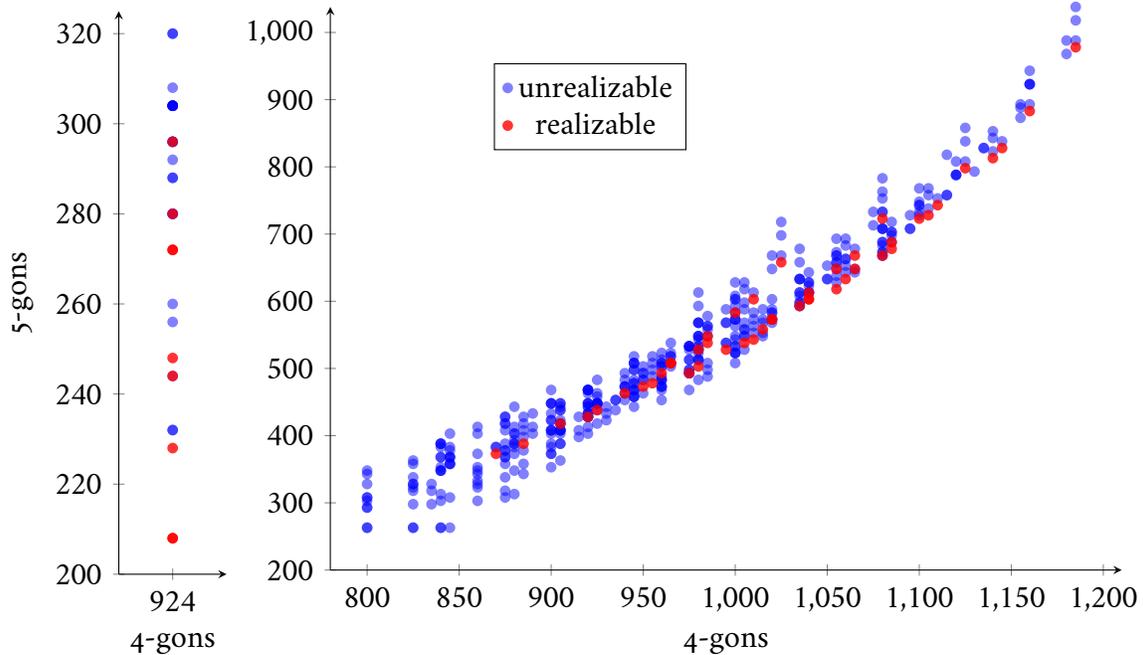

We also examined the number of $4$-gons and $5$-gons in the solutions. All 66 solutions
of the $4$-fold symmetry have 924 $4$-gons, while the number of $5$-gons ranges between 208 and 320. 
For the $5$-fold symmetry solutions, the number of $4$-gons ranges from 800 to 1185, while the number of
$5$-gons range from 263 to 1038.
Figure~\ref{fig:scatter} illustrates the data. Note that assignments are more likely to be realizable if they 
have a relatively high number of $4$-gons and a relatively low number of $5$-gons.

\subsection{Avoiding $7$-Gons}

We also sought a symmetric solution for 32 points without a $7$-gon. On 32 points, only $7$-gon-free $s$-fold rotational symmetries could exist with $s \in \{1,2,4\}$. We focused our experiments on $s=4$ as that symmetry reduces the search space the most. The resulting formula is easy to satisfy, but the initial solutions that we obtained were hard to realize.
We therefore decided to enumerate all solutions on a supercomputer, which took about 1 CPU year. The number of non-isomorphic configurations is staggering: 310\,187\,713. We attempted to realize several of them using heuristics to determine which ones would be more likely to be realizable. During those experiments, we observed that the outer 28 points can frequently be realized (so only the inner 4 points not), while the inner 12 points were never realizable. Afterwards we observed that all 310 million solutions have only 6 different configurations to the inner 12 points and none of those are realizable. Thus, there does not exist a realizable 4-fold symmetry on 32 points without a $7$-gon. 

\subsection{Avoiding Balance}
\begin{figure}[H]
\centering
  \begin{tikzpicture}[dot/.style={inner sep=1pt, circle, minimum size=1.5mm, thick, scale=0.7}]
    \begin{axis}[
      unit vector ratio=1 1 1,
      xmin=-5.5,xmax=4.,
      ymin=-4.5,ymax=5.5,
      axis lines=none,
      xticklabel=\empty,
      yticklabel=\empty ]
\node[dot, fill=blue!100!red] (1) at (axis cs:-5.196152422706632, -1) {\lab{1}};
\node[dot, fill=blue!60!red] (2) at (axis cs:3.497173145439157, -1.384939661076581) {\lab{2}};
\node[dot, fill=blue!80!red] (3) at (axis cs:2.598076211353317, -2.4999999999999982) {\lab{3}};
\node[dot, fill=blue!40!red] (4) at (axis cs:1.7320508075688772, -1) {\lab{4}};
\node[dot, fill=blue!10!red] (5) at (axis cs:0.8660254037844384, 0.5000000000000003) {\lab{5}};
\node[dot, fill=blue!70!red] (6) at (axis cs:3.148267213139839, -1.8176529229871585) {\lab{6}};
\node[dot, fill=blue!100!red] (8) at (axis cs:1.7320508075688759, 5.0) {\lab{8}};
\node[dot, fill=blue!60!red] (9) at (axis cs:-2.9479795019205075, -2.3361709548447513) {\lab{9}};
\node[dot, fill=blue!80!red] (10) at (axis cs:-3.4641016151377544, -1) {\lab{10}};
\node[dot, fill=blue!40!red] (11) at (axis cs:-1.7320508075688772, -1) {\lab{11}};
\node[dot, fill=blue!10!red] (12) at (axis cs:0, -1) {\lab{12}};
\node[dot, fill=blue!70!red] (13) at (axis cs:-3.148267213139838, -1.8176529229871599) {\lab{13}};
\node[dot, fill=blue!100!red] (15) at (axis cs:3.464101615137756, -3.9999999999999982) {\lab{15}};
\node[dot, fill=blue!60!red] (16) at (axis cs:-0.5491936435186506, 3.721110615921331) {\lab{16}};
\node[dot, fill=blue!80!red] (17) at (axis cs:0.8660254037844378, 3.5) {\lab{17}};
\node[dot, fill=blue!40!red] (18) at (axis cs:0, 2) {\lab{18}};
\node[dot, fill=blue!10!red] (19) at (axis cs:-0.8660254037844387, 0.4999999999999998) {\lab{19}};
\node[dot, fill=blue!70!red] (20) at (axis cs:-0.00000019984014443252818, 3.635305845974317) {\lab{20}};

  \draw (1) -- (10) -- (11) -- (12) -- (4);
  \draw (8) -- (17) -- (18) -- (19) -- (11);
  \draw (15) -- (3) -- (4) -- (5) -- (18);

\node[dot, fill=blue!25!red] (7) at (axis cs:-1.4566615990962988, -0.3046023798138631) {\lab{7}};
\node[dot, fill=blue!25!red] (14) at (axis cs:0.4645374005761473, 1.4138071394415899) {\lab{14}};
\node[dot, fill=blue!25!red] (21) at (axis cs:0.9921241985201515, -1.1092047596277261) {\lab{21}};

  \draw[line width=0.05mm] (5) -- (7) -- (10);
  \draw[line width=0.05mm] (12) -- (14) -- (17);
  \draw[line width=0.05mm] (19) -- (21) -- (3);

  \draw[line width=0.05mm] (2) -- (4) -- (7);
  \draw[line width=0.05mm] (2) -- (21) -- (12);

  \draw[line width=0.05mm] (9) -- (11) -- (14);
  \draw[line width=0.05mm] (9) -- (7) -- (19);  

  \draw[line width=0.05mm] (16) -- (18) -- (21);
  \draw[line width=0.05mm] (16) -- (14) -- (5);  
  
  \draw[line width=0.05mm]  (6) -- (21) -- (7);
  \draw[line width=0.05mm]  (6) -- (4) -- (19);
  
  \draw[line width=0.05mm]  (13) -- (7) -- (14);
  \draw[line width=0.05mm]  (13) -- (11) -- (5);

  \draw[line width=0.05mm]  (20) -- (14) -- (21);
  \draw[line width=0.05mm]  (20) -- (18) -- (12);
  
  \draw[line width=0.05mm]  (2) -- (6) -- (3);
  \draw[line width=0.05mm]  (9) -- (13) -- (10);
  \draw[line width=0.05mm]  (16) -- (20) -- (17);

%
%
%
%
%

    \end{axis}
  \end{tikzpicture}
\caption{A $2$-everywhere-unbalanced construction on $21$ points.}
\label{fig:unbalanced}
\end{figure}
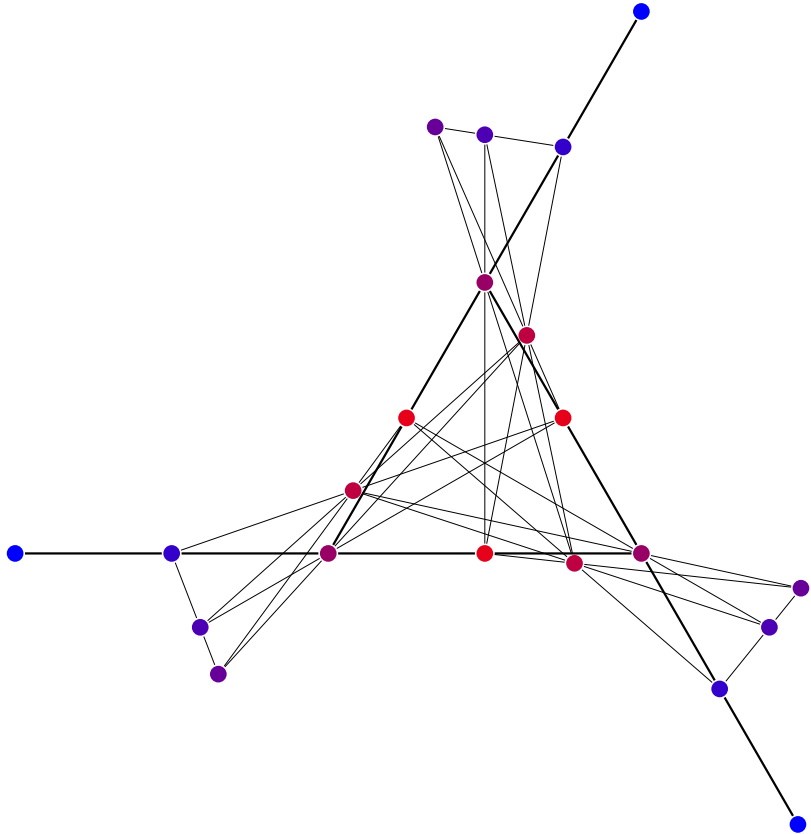

It is known from prior constructions \cite{ResearchProblems2002,conlon2025unbalancedconfigurations} that there exist infinitely many point sets with an even number of points having imbalance 2, as well as a set of 23 points having imbalance 2. However, it was unknown if these constructions were minimal. So we explored the question: What is the smallest even (odd) number of points needed to achieve an imbalance of 2? Utilizing our encoding, we were first able to show that the 12-point construction in \cite{ResearchProblems2002} is indeed minimal, as a smaller number of points produces encodings that are UNSAT. Similarly, in the odd case, we were able to refute the existence of such point sets having $\leq 19$ points. Furthermore, by searching for solutions with a $3$-fold (combinatorial) rotational symmetry, we were able to find \emph{satisfying and realizable} solutions with 21 points (\Cref{fig:unbalanced}), thereby completely answering the minimality questions. 

\section{Conclusion}

We have presented a systematic approach to obtain symmetric constructions for discrete geometric problems, and along the way, a highly performant local-search solver for the realizability problem. 
Our results show further evidence of an interesting pattern highlighted in the introduction: while general constructions for problems in Ramsey theory are often asymmetric, it turns out that there are symmetric solutions for small values of the parameters.
A natural direction of research, both for computer scientists and mathematicians, is to try to uncover general constructions from the symmetric solutions that we can find for small instances.
In particular, for the Erd\H{o}s-Szekeres problem, we suspect that there are symmetric constructions for all $k \geq 7$, and proving this could represent progress toward the conjectured bound $g(k) = 2^{k-2} + 1$, since it might be easier to prove optimality for more structured solutions.
In the case of the everywhere-unbalanced-points problem, our future research will focus on finding a $3$-EU set, for which we aim to leverage the reduction in the number of variables achieved by forcing an ansatz symmetry.
Further problems in this line of work include finding a realization of a $2$-fold symmetric construction of $32$ points without a convex $7$-gon, and designing an efficient solver that can handle collinearity constraints for the realizability problem.

\bibliographystyle{plain}
\bibliography{references}
%


\appendix

\section{Realizations of 4-Fold Symmetric Configurations}

Below are the nine non-isomorphic realizable configurations on 16 points with a 4-fold rotational symmetry.
The lines are added to make the symmetry more clearly visible and to facilitate comparison of the different configurations. 
Note that in the top three plots, the two outer layers are very close to each other, thereby resulting in overlapping points. 

\newcommand{\fourplot}[1]{
    \begin{subfigure}{0.31\textwidth}
        \centering
        \fbox{\includegraphics[height=120pt]{#1}}
    \end{subfigure}
}

\begin{figure}[h!]
\fourplot{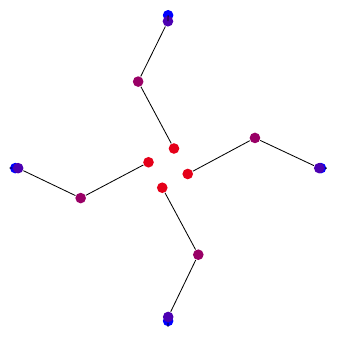} \hfill
\fourplot{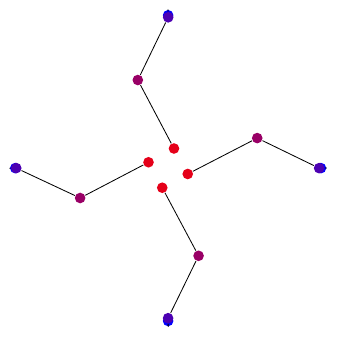} \hfill
\fourplot{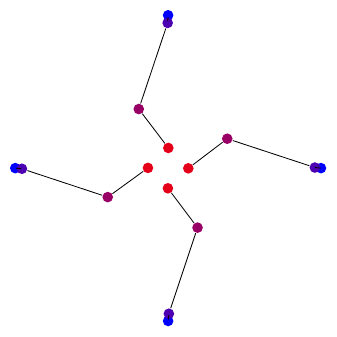} 
\end{figure}

\begin{figure}[h!]
\fourplot{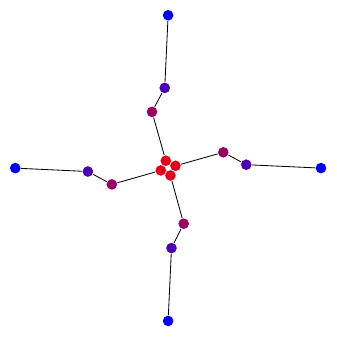} \hfill
\fourplot{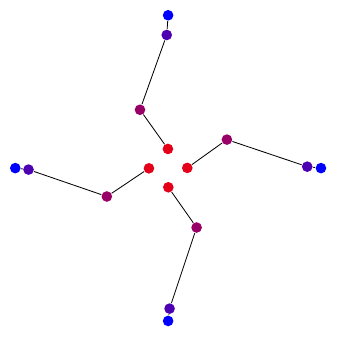} \hfill
\fourplot{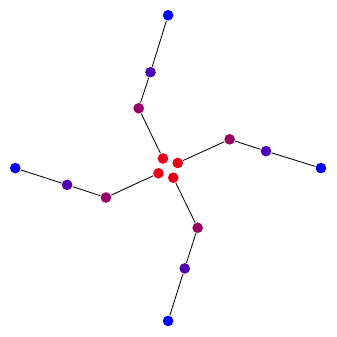} 
\end{figure}

\begin{figure}[h!]
\fourplot{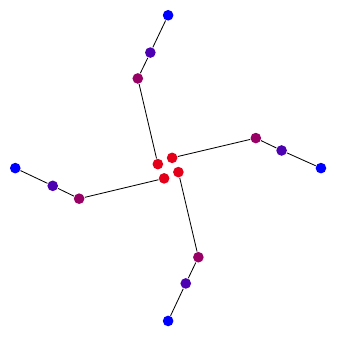} \hfill
\fourplot{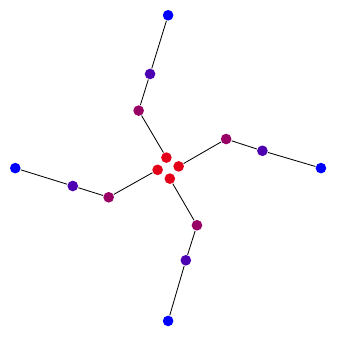} \hfill
\fourplot{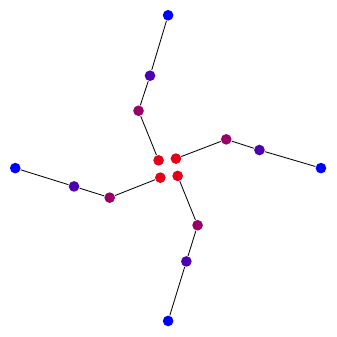} 
\end{figure}

\newpage

\section{Realizations of 5-Fold Symmetric Configurations}

Below are the 46 non-isomorphic realizable configurations on 16 points with a 5-fold rotational symmetry.
The lines are added to make the symmetry more clearly visible and help compare the different configurations. 

\newcommand{\fiveplot}[1]{
    \begin{subfigure}{0.23\textwidth}
        \centering
        \fbox{\includegraphics[height=100pt]{#1}}
    \end{subfigure}
}

\begin{figure}[h!]

\fiveplot{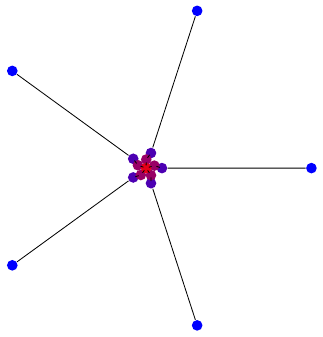} \hfill
\fiveplot{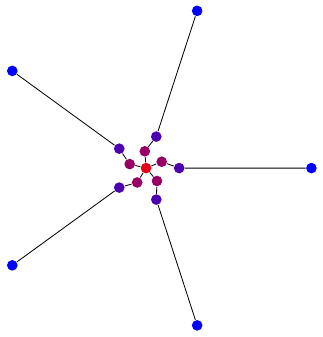} \hfill
\fiveplot{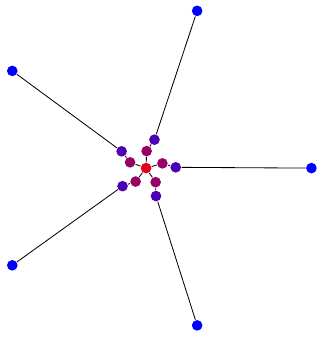} \hfill
\fiveplot{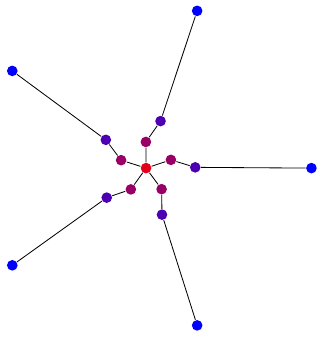}
\end{figure}

\begin{figure}[h!]
\fiveplot{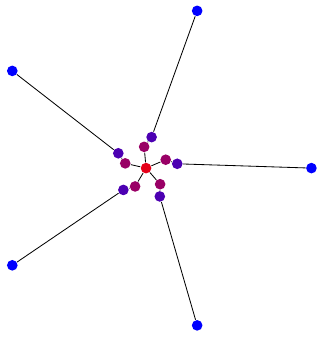} \hfill
\fiveplot{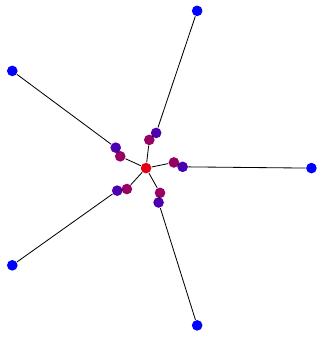} \hfill
\fiveplot{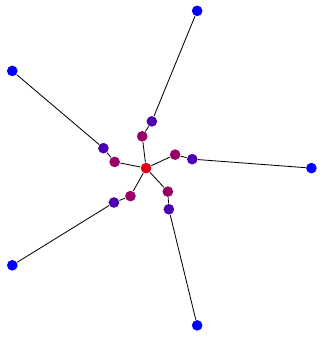} \hfill
\fiveplot{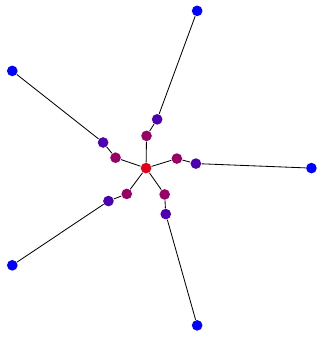}
\end{figure}

\begin{figure}[h!]
\fiveplot{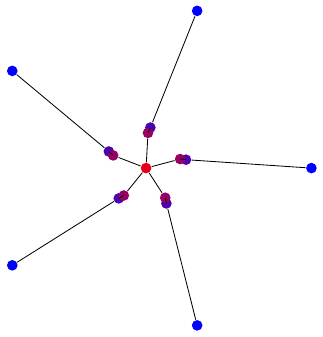} \hfill
\fiveplot{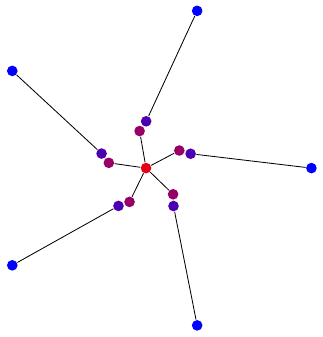} \hfill
\fiveplot{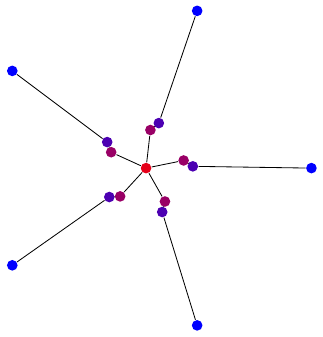} \hfill
\fiveplot{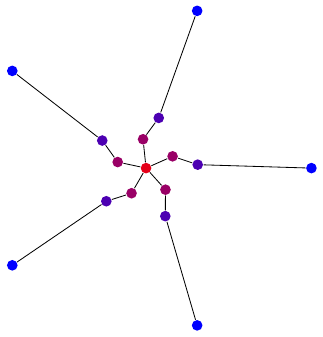}
\end{figure}

\begin{figure}[h!]
\fiveplot{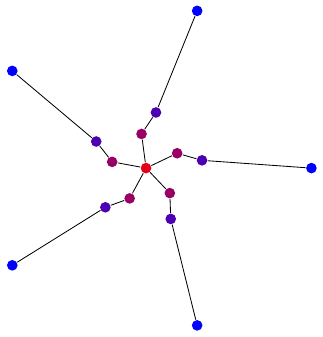} \hfill
\fiveplot{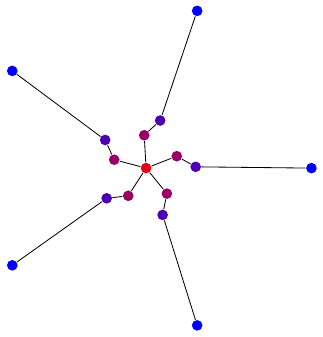} \hfill
\fiveplot{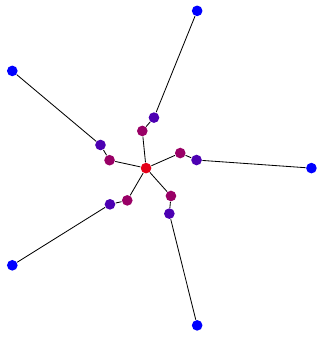} \hfill
\fiveplot{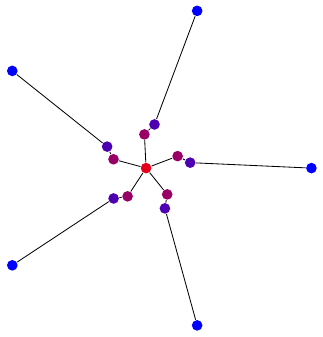}
\end{figure}

\begin{figure}[h!]
\fiveplot{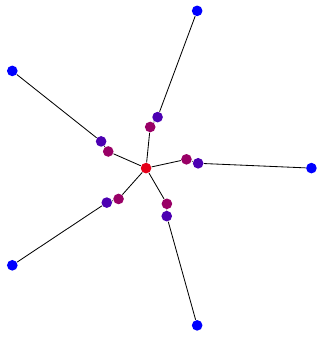} \hfill
\fiveplot{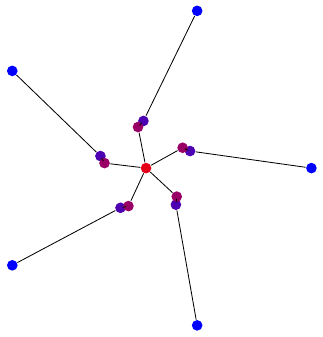} \hfill
\fiveplot{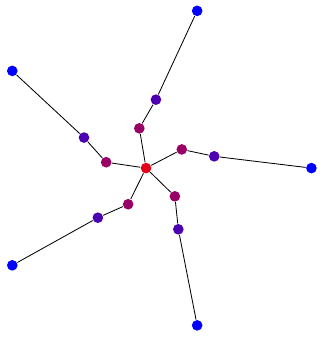} \hfill
\fiveplot{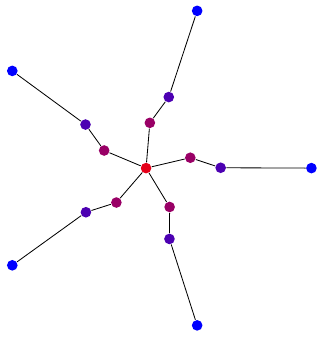}
\end{figure}

\begin{figure}[h!]
\fiveplot{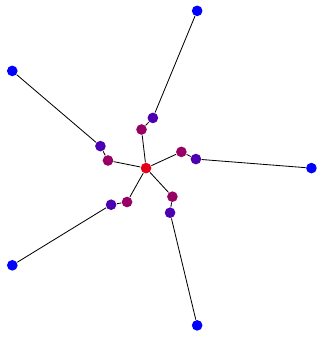} \hfill
\fiveplot{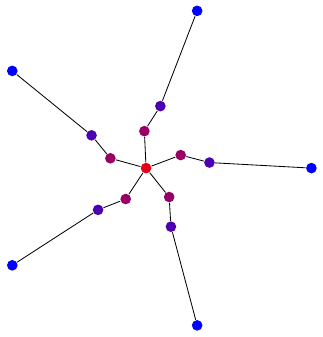} \hfill
\fiveplot{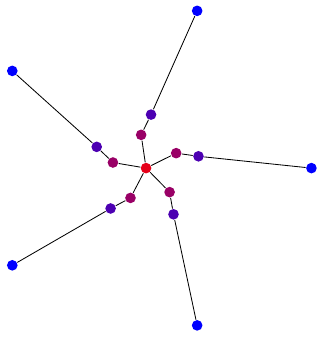} \hfill
\fiveplot{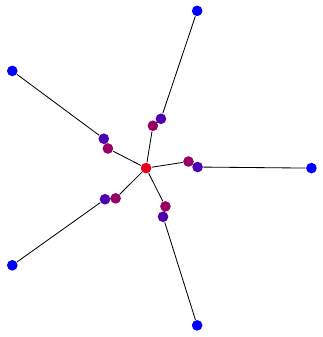} \hfill
\end{figure}


\begin{figure}[h!]
\fiveplot{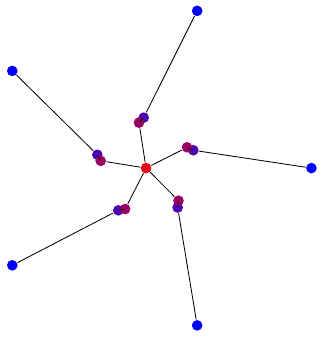} \hfill
\fiveplot{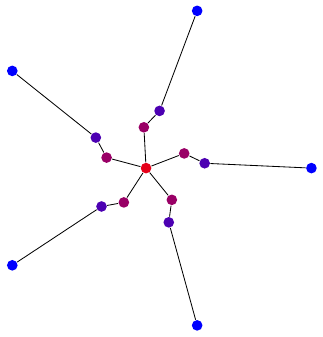} \hfill
\fiveplot{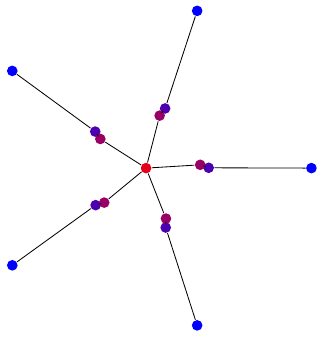} \hfill
\fiveplot{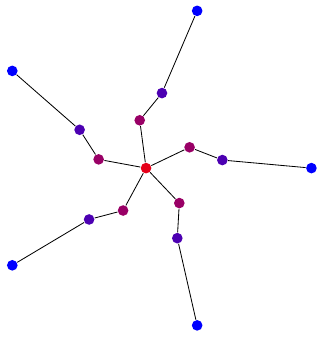}
\end{figure}

\begin{figure}[h!]
\fiveplot{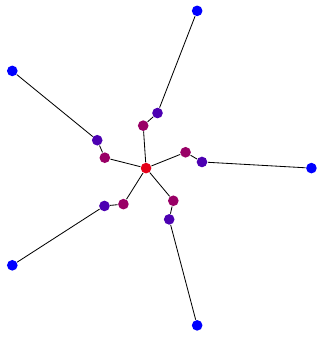} \hfill
\fiveplot{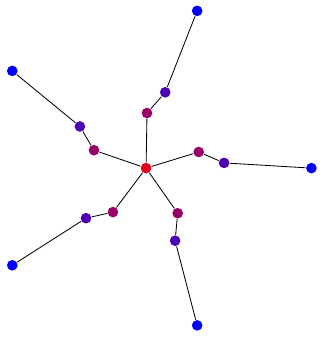} \hfill
\fiveplot{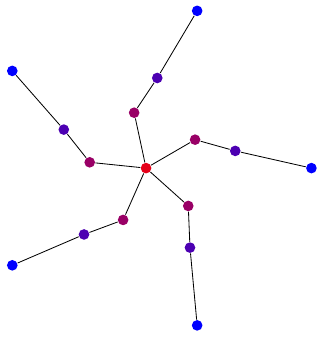} \hfill
\fiveplot{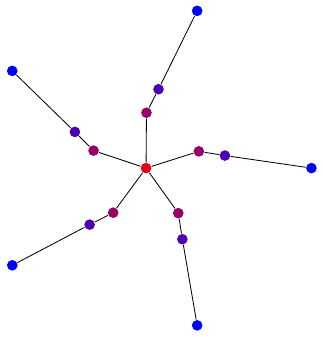}
\end{figure}

\begin{figure}[h!]
\fiveplot{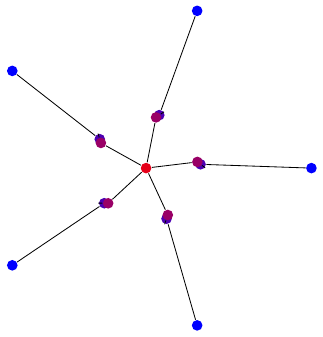} \hfill
\fiveplot{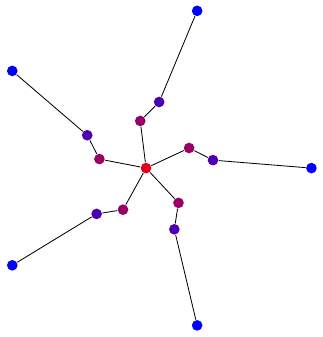} \hfill
\fiveplot{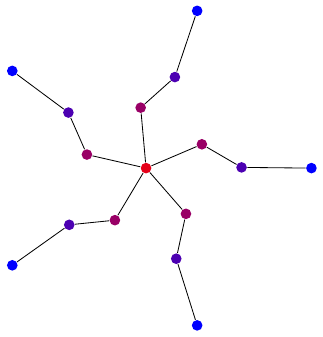} \hfill
\fiveplot{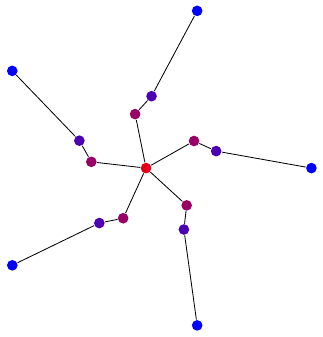}
\end{figure}

\begin{figure}[h!]
\fiveplot{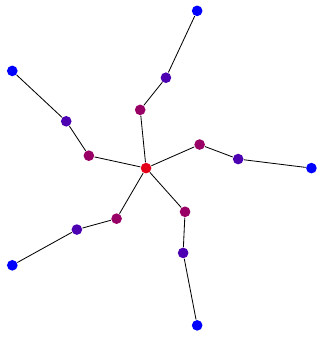} \hfill
\fiveplot{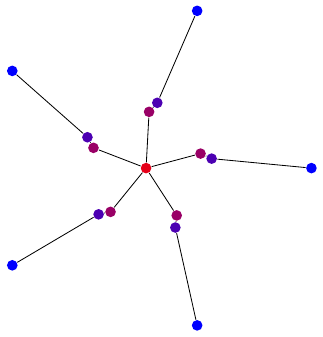} \hfill
\fiveplot{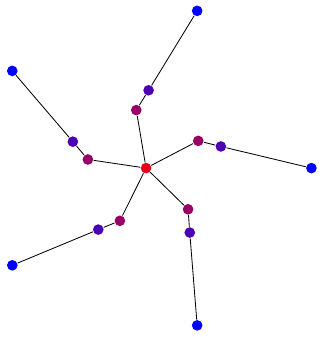} \hfill
\fiveplot{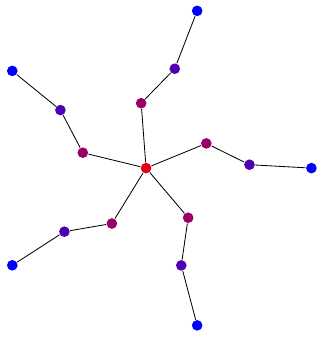}
\end{figure}


\begin{figure}[h!]
\fiveplot{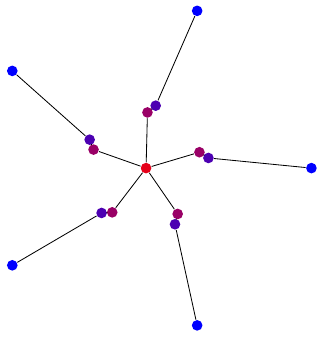} \hfill
\fiveplot{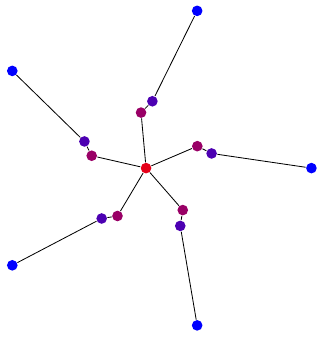} \hfill
\fiveplot{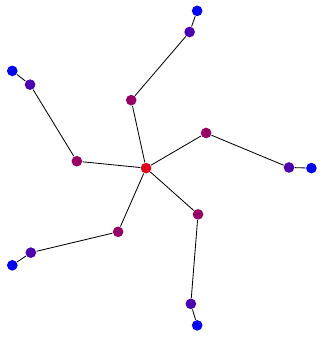} \hfill
\fiveplot{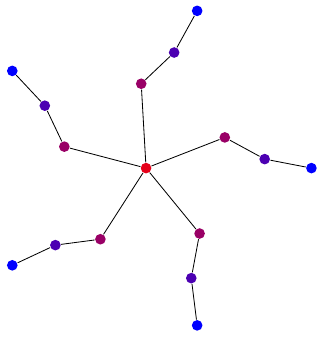}
\end{figure}

\begin{figure}[h!]
\fiveplot{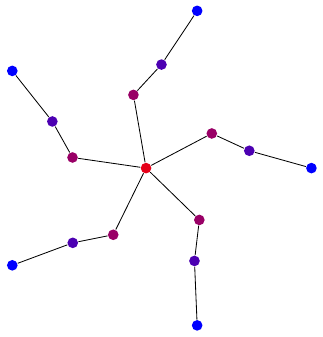}
\fiveplot{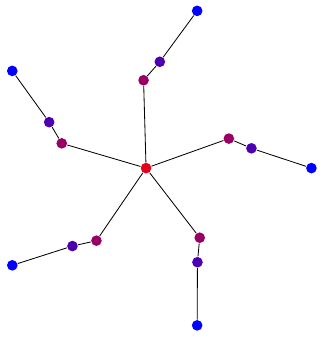}
\end{figure}

\newpage

\section{Explicit coordinates}\label{sec:exact}

We present, for verification purposes, the coordinates of the points in the 4-fold and 5-fold symmetric configurations presented in~\Cref{fig:16-4fold,fig:16-5fold}.
The coordinates for~\Cref{fig:16-4fold} are presented explicitly in~\Cref{tab:16-4fold-coords}, whereas the coordinates for~\Cref{fig:16-5fold} are better described by specifying some rational points and leaving the rest implicit. 
Concretely, if we denote again $\rho_{2\pi/5}(x, y) = (x \cos(2\pi/5) - y \sin(2\pi/5), x \sin(2\pi/5) + y \cos(2\pi/5))$, and denote by $\rho_{2\pi/5}^(k)$ the $k$-fold application of $\rho_{2\pi/5}$, then~\Cref{tab:16-5fold-coords} describes the coordinates of the points in~\Cref{fig:16-5fold}.

\begin{table}[h]
\caption{The coordinates of the points in~\Cref{fig:16-4fold}.}\label{tab:16-4fold-coords}
\begin{tabular}{ll}
 1: & $(-30, 0)$\\
 2: & $(0, -30)$\\
 3: & $(30, 0)$\\
 4: & $(0, 30)$\\
\end{tabular}
\hfil
\begin{tabular}{ll}
 5: & $\left(-20, -\frac{7}{2}\right)$\\
 6: & $\left(\frac{7}{2}, -20\right)$\\
 7: & $\left(20, \frac{7}{2}\right)$\\
 8: & $\left(-\frac{7}{2}, 20\right)$\\
\end{tabular}
\hfil
\begin{tabular}{ll}
 9: & $(-13, -6)$\\
 10: & $(6, -13)$\\
 11: & $(13, 6)$\\
 12: & $(-6, 13)$\\
\end{tabular}
\hfil
\begin{tabular}{ll}
 13: & $\left(-\frac{19}{10}, -\frac{6}{5}\right)$\\
 14: & $\left(\frac{6}{5}, -\frac{19}{10}\right)$\\
 15: & $\left(\frac{19}{10}, \frac{6}{5}\right)$\\
 16: & $\left(-\frac{6}{5}, \frac{19}{10}\right)$\\
\end{tabular}
\end{table}

\begin{table}[h]
\caption{The coordinates of the points in~\Cref{fig:16-5fold}.}\label{tab:16-5fold-coords}
\begin{tabular}{ll}
    \vspace{0.2cm}
 1: & $\rho_{2\pi/5}^{(4)}(-12, -17)$\\
 \vspace{0.2cm}
 2: & $(-12, -17)$\\
  \vspace{0.2cm}
 3: & $\rho_{2\pi/5}^{(1)}(-12, -17)$\\
  \vspace{0.2cm}
 4: & $\rho_{2\pi/5}^{(2)}(-12, -17)$\\
  \vspace{0.2cm}
 5: & $\rho_{2\pi/5}^{(3)}(-12, -17)$\\
\end{tabular}
\hfil
\begin{tabular}{ll}
    \vspace{0.2cm}
 6: & $(-15, 2)$\\
   \vspace{0.2cm}
 7: & $\rho_{2\pi/5}^{(1)}(-15, 2 )$\\
   \vspace{0.2cm}
 8: & $\rho_{2\pi/5}^{(2)}(-15, 2 )$\\
   \vspace{0.2cm}
 9: & $\rho_{2\pi/5}^{(3)}(-15, 2 )$\\
   \vspace{0.2cm}
 10: & $\rho_{2\pi/5}^{(4)}(-15, 2 )$\\
\end{tabular}
\hfil
\begin{tabular}{ll}
      \vspace{0.2cm}
 11: & $(-13, 0)$\\
   \vspace{0.2cm}
 12: & $\rho_{2\pi/5}^{(1)}(-13, 0)$\\
   \vspace{0.2cm}
 13: & $\rho_{2\pi/5}^{(2)}(-13, 0)$\\
   \vspace{0.2cm}
 14: & $\rho_{2\pi/5}^{(3)}(-13,0)$\\
   \vspace{0.2cm}
 15: & $\rho_{2\pi/5}^{(4)}(-13, 0)$\\
\end{tabular}
\hfil
\begin{tabular}{ll}
 16: & $(0, 0)$\\
\end{tabular}
\end{table}

\begin{figure}[h]
\centering
  \begin{tikzpicture}[dot/.style={inner sep=1pt, circle, minimum size=1.5mm, thick, scale=0.7}]
    \begin{axis}[
      unit vector ratio=1 1 1,
      xmin=-5.5,xmax=6.25,
      ymin=-4.5,ymax=6.25,
      axis lines=none,
      xticklabel=\empty,
      yticklabel=\empty ]
\node[dot, fill=blue!100!red] (1) at (axis cs:{-3*\toint}, -1) {\lab{1}};
\node[dot, fill=blue!60!red] (2) at (axis cs:{36/11*\toint}, -20/11) {\lab{2}};
\node[dot, fill=blue!80!red] (3) at (axis cs:{3/2*\toint}, -5/2) {\lab{3}};
\node[dot, fill=blue!40!red] (4) at (axis cs:{1*\toint}, -1) {\lab{4}};
\node[dot, fill=blue!10!red] (5) at (axis cs:{1/2*\toint}, 1/2) {\lab{5}};
\node[dot, fill=blue!70!red] (6) at (axis cs:{961/520*\toint}, -961/520) {\lab{6}};
\node[dot, fill=blue!100!red] (8) at (axis cs:{1*\toint}, 5) {\lab{8}};
\node[dot, fill=blue!60!red] (9) at (axis cs:{-28/11*\toint}, -4) {\lab{9}};
\node[dot, fill=blue!80!red] (10) at (axis cs:{-2*\toint}, -1) {\lab{10}};
\node[dot, fill=blue!40!red] (11) at (axis cs:{-1*\toint}, -1) {\lab{11}};
\node[dot, fill=blue!10!red] (12) at (axis cs:{0*\toint}, -1) {\lab{12}};
\node[dot, fill=blue!70!red] (13) at (axis cs:{-961/520*\toint}, -961/520) {\lab{13}};
\node[dot, fill=blue!100!red] (15) at (axis cs:{2*\toint}, -4) {\lab{15}};
\node[dot, fill=blue!60!red] (16) at (axis cs:{-8/11*\toint}, 64/11) {\lab{16}};
\node[dot, fill=blue!80!red] (17) at (axis cs:{1/2*\toint}, 7/2) {\lab{17}};
\node[dot, fill=blue!40!red] (18) at (axis cs:{0*\toint}, 2) {\lab{18}};
\node[dot, fill=blue!10!red] (19) at (axis cs:{-1/2*\toint}, 1/2) {\lab{19}};
\node[dot, fill=blue!70!red] (20) at (axis cs:{0*\toint}, 961/260) {\lab{20}};      
      
  \draw (1) -- (10) -- (11) -- (12) -- (4);
  \draw (8) -- (17) -- (18) -- (19) -- (11);
  \draw (15) -- (3) -- (4) -- (5) -- (18);

\node[dot, fill=blue!25!red] (7) at (axis cs:{-7/8*\toint}, -13/40) {\lab{7}};
\node[dot, fill=blue!25!red] (14) at (axis cs:{11/40*\toint}, 59/40) {\lab{14}};
\node[dot, fill=blue!25!red] (21) at (axis cs:{3/5*\toint}, -23/20) {\lab{21}};
  
  \draw[line width=0.05mm] (5) -- (7) -- (10);
  \draw[line width=0.05mm] (12) -- (14) -- (17);
  \draw[line width=0.05mm] (19) -- (21) -- (3);

  \draw[line width=0.05mm] (2) -- (4) -- (7);
  \draw[line width=0.05mm] (2) -- (21) -- (12);

  \draw[line width=0.05mm] (9) -- (11) -- (14);
  \draw[line width=0.05mm] (9) -- (7) -- (19);  

  \draw[line width=0.05mm] (16) -- (18) -- (21);
  \draw[line width=0.05mm] (16) -- (14) -- (5);  
  
  \draw[line width=0.05mm]  (6) -- (21) -- (7);
  \draw[line width=0.05mm]  (6) -- (4) -- (19);
  
  \draw[line width=0.05mm]  (13) -- (7) -- (14);
  \draw[line width=0.05mm]  (13) -- (11) -- (5);

  \draw[line width=0.05mm]  (20) -- (14) -- (21);
  \draw[line width=0.05mm]  (20) -- (18) -- (12);
  

    \end{axis}
  \end{tikzpicture}
\caption{A $21$-point $2$-everywhere-unbalanced configuration with simple coordinates.}
\label{fig:simple}
\end{figure}
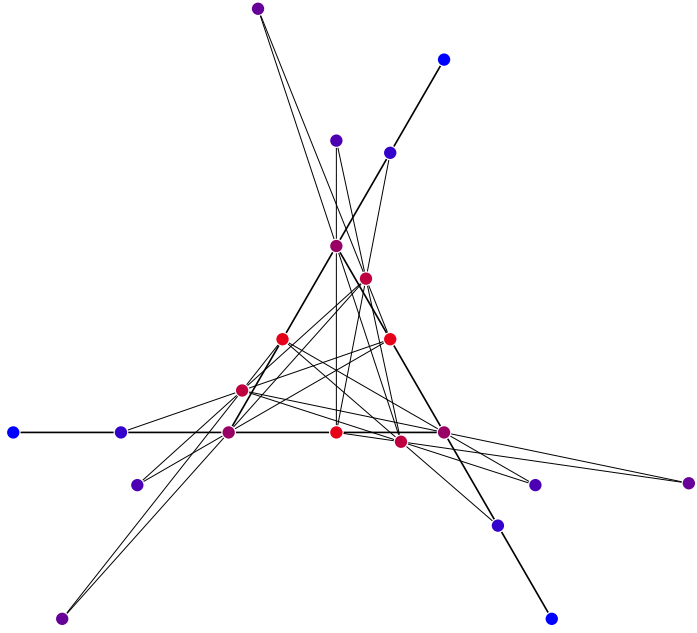

\begin{table}[h]
\caption{The coordinates of the points in Figure~\ref{fig:simple}.}
\begin{tabular}{r@{~}l}
 1: & $(-3\sqrt{3}, -1)$\\
 2: & $(36/11\sqrt{3}, -20/11)$\\
 3: & $(3/2\sqrt{3}, -5/2)$\\
 4: & $(\sqrt{3}, -1)$\\
 5: & $(1/2\sqrt{3}, 1/2)$\\
 6: & $(961/520\sqrt{3}, -961/520)$\\
 7: & $(-7/8\sqrt{3}, -13/40)$
\end{tabular}
\hfil
\begin{tabular}{r@{~}l}
 8: & $(\sqrt{3}, 5)$\\
 9: & $(-28/11\sqrt{3}, -4)$\\
 10: & $(-2\sqrt{3}, -1)$\\
 11: & $(-\sqrt{3}, -1)$\\
 12: & $(0, -1)$\\
 13: & $(-961/520\sqrt{3}, -961/520)$\\
 14: & $(11/40\sqrt{3}, 59/40)$
\end{tabular}
\hfil
\begin{tabular}{r@{~}l}
 15: & $(2\sqrt{3}, -4)$\\
 16: & $(-8/11\sqrt{3}, 64/11)$\\
 17: & $(1/2\sqrt{3}, 7/2)$\\
 18: & $(0, 2)$\\
 19: & $(-1/2\sqrt{3}, 1/2)$\\
 20: & $(0, 961/260)$\\
 21: & $(3/5\sqrt{3}, -23/20)$
\end{tabular}
\end{table}

\section{Proofs for the dynamic-ordering axioms}\label{sec:dynamic_axiom1_proof}

\begin{proof}[Proof of~\Cref{prop:dynamic_axiom1}]

    We first prove that~\Cref{eq:dynamic_axiom1} is satisfied. Up to relabeling, let $\{p_1, p_2, p_3, p_4\}$ be $4$ points from $S$ such that the ordering requirements $\prec_{1, 2} \land \prec_{1, 3} \land \prec_{1, 4}$ are satisfied.
    By translating if necessary, we may further assume that $p_1 = (0, 0)$ is the origin (note that this does not affect the relative ordering of the $x$-coordinates nor the orientations). 
    Then, we will think of the condition $\varA_{1,2,3} \lor \overline{\varA_{1, 2, 4}} \lor \varA_{1, 3, 4}$ as $(\overline{\varA_{1,2,3}} \land \overline{\varA_{1,3,4}}) \rightarrow \overline{\varA_{1,2,4}}$.
    Thus, we assume that $\overline{\varA_{1, 2, 3}}, \overline{\varA_{1, 3, 4}}$ hold, and it remains to prove that $\overline{\varA_{1, 2, 4}}$ holds as well. 
    By the construction of the assignment $\tau$, we have that $\sigma(p_1, p_2, p_3) = -1$ and $\sigma(p_1, p_3, p_4) = -1$.
    Recall that the orientation $\sigma(p, q, r)$ is defined according to the sign of the determinant (or cross-product) relating the points:
    \begin{equation*}
       \text{sign} \det \begin{pmatrix} p_x & q_x & r_x \\ p_y & q_y & r_y \\ 1 & 1 & 1 \end{pmatrix} = \begin{cases}
         -1 & \text{if } (p_x - q_x)(p_y - r_y)  < (p_y - q_y)(p_x - r_x),\\
          0 & \text{if } (p_x - q_x)(p_y - r_y)  = (p_y - q_y)(p_x - r_x),\\
           1 & \text{if } (p_x - q_x)(p_y - r_y)  > (p_y - q_y)(p_x - r_x).
            \end{cases}
    \end{equation*}

    Therefore, from $\sigma(p_1, p_2, p_3) = -1$ we have 
    \begin{equation}
        x_2y_3 < x_3y_2,
    \end{equation}
    and from $\sigma(p_1, p_3, p_4) = -1$ we have 
    \begin{equation}
        \label{eq:proof_ineq_2}
        x_3y_4 < x_4y_3.
    \end{equation}
    Since $\prec_{1, 2}$ holds, we have $x_1 < x_2$ and thus we can multiply~\Cref{eq:proof_ineq_2} by $x_2$ without flipping the inequality, obtaining
    \begin{align*}
        x_2x_3y_4 &< x_2x_4y_3 = x_4(x_2y_3) < x_4(x_3y_2).
    \end{align*}
    By dividing the left- and right-most terms by $x_3$, which is positive since $\prec_{1, 3}$ holds, we have $x_2y_4 < x_4y_2$. This directly implies that $\sigma(p_1, p_2, p_4) = -1$, and thus $\overline{\varA_{1, 2, 4}}$ holds, satisfying~\Cref{eq:dynamic_axiom1}.

   The proof for~\Cref{eq:dynamic_axiom2} is similar, but this time setting $p_3 = (0, 0)$ for convenience. Here we assume $\overline{\varA_{1, 2, 3}}$ and $\overline{\varA_{2, 3, 4}}$, from where we have 
    \begin{align}
    x_4y_2 &< x_2y_4,\label{eq:proof42}\\
    x_1y_2 &< x_2y_1\label{eq:proof12}.
    \end{align}
    We moreover have, due to the ordering assumption, that $x_1, x_2 < 0$ and $x_4 > 0$, and we want to show that $\overline{\varA_{1, 3, 4}}$ holds. By the equivalence discussed in~\Cref{subsec:cc_systems}, this is the same as proving that $\overline{\varA_{3, 4, 1}}$ holds, which we do next. 
    Indeed, since $x_1 < 0$, multiplying~\Cref{eq:proof42} by $x_1$ gives
     \begin{equation}x_1x_4y_2 > x_1x_2y_4,\label{eq:proof142}\end{equation} and as $x_4 > 0$, multiplying~\Cref{eq:proof12} by $x_4$ gives
     \begin{equation} x_4x_1y_2 < x_4x_2y_1.\label{eq:proof412}\end{equation}
Then, transitivity over~\Cref{eq:proof142,eq:proof412} gives
\[ 
    x_1x_2y_4 < x_4x_2y_1, 
\]
which dividing by $x_2$ (recalling that $x_2 < 0$) gives $x_1y_4 > x_4y_1$,
    which is equivalent to $\overline{\varA_{3, 4, 1}}$, as desired. 

\end{proof}

\begin{proof}[Proof of~\Cref{prop:dynamic_axiom2}]
    
    We first prove that it suffices to show it for $n = 5$. Indeed, suppose that for some $n > 5$ there is an assignment $\tau$ of the $\varA_{i, j, k}$ variables, and $\theta$ for the $\prec_{i, j}$ variables, that satisfies 
    the dynamic-ordering axioms but not the CC axioms. Then, in particular, there is a CC-axiom clause $C$ that is not satisfied by $\tau$, which involves a set $S$ of at most $5$ indices (see~\Cref{subsec:cc_systems}). Thus, the restriction of $\theta \cup \tau$ to variables whose indices are contained in $S$, which we denote by $(\theta \cup \tau)_{|S}$, holds $(\theta \cup \tau)_{|S} \nvDash C$. On the other hand, if we denote by $D_S$ the set of clauses of the dynamic-ordering axioms that involve only variables whose indices are in $S$, we have $(\theta \cup \tau)_{|S} \vDash D_S$. By considering the mapping $f: S \to \{1, \ldots, |S|\}$ that maps the $i$-th largest index in $S$ to $i$, and letting $C^f$ (resp. ${D^f_S}$) be clauses obtained by replacing each index $j$ appearing in some variable in $C$ (resp. $D_S$) with its image $f(j)$, then we have that $(\theta \cup \tau)_{|f(S)} \vDash {D^f_S}$ and $(\theta \cup \tau)_{|f(S)} \nvDash C^f$, which would consitute a counterexample with $n \leq 5$.
    For $n = 5$ (which trivially implies the cases with $n < 5$), the question reduces to a finite computation, which we carry out as a SAT problem. We start by we creating a formula $\textsf{DOA}(5)$, consisting of the dynamic-ordering axioms for $n = 5$, and a formula $\textsf{CCA}(5)$, consisting of the CC axioms for $n = 5$. 
    We then use a Tseitin transformation to negate the CC axioms: for each clause $C \in \textsf{CCA}(5)$, we introduce a new variable $y_C$ and add a clause $\overline{\ell} \lor \overline{y_C}$ for each literal $\ell$ in $C$, thus ensuring that if clause $C$ is satisfied by any of its literals, then $y_C$ is false. Therefore, given that $\textsf{DOA}(5)$ is clearly satisfiable, the formula
    \[
       \Psi := \textsf{DOA}(5) \land \bigwedge_{C \in \textsf{CCA}(5)} \bigwedge_{\ell \in C} \left(\overline{\ell} \lor \overline{y_C}\right) \land \left(\bigvee_{C \in \textsf{CCA}(5)} y_C\right)
    \]
    is satisfiable if and only if there is an assignment $\tau$ of the $\varA_{i, j, k}$ variables and $\theta$ for the $\prec_{i, j}$ variables for which $(\tau \cup \theta)$ satisfies the dynamic-ordering axioms, but not the CC axioms.
    We conclude by running a SAT solver on $\Psi$, and finding that $\Psi$ is unsatisfiable, which proves that no such assignment exists. The code for generating the formulas is available in the repository \url{github.com/bsubercaseaux/automatic-symmetries}.
\end{proof}

\section{Pseudocode for~\localizer}\label{sec:local_realizer_pseudocode}

Some auxiliary functions used by~\localizer~are presented in~\Cref{alg:auxiliary}. The main loop of each thread is presented in~\Cref{alg:localizer_thread}. 
About~\Cref{alg:auxiliary}, it is worth clarifying that an important point is that when moving a given point $p_i$, only the orientation constraints involving $p_i$ can change whether they are satisfied or not, and thus we can do an $O(n^2)$ evaluation, instead of the general $O(n^3)$ evaluation.

\begin{algorithm}
    \caption{Auxiliary functions for~\localizer.}
    \label{alg:auxiliary}
    \begin{algorithmic}[1]
    \Require An orientation assignment $\tau : \binom{n}{3} \to \{-1, 1\}$.
    
    \Function{Eval}{$P, \tau$}
        \State $u \gets 0$ \Comment{Number of unsat constraints}
        \State $F \gets [0, 0, \ldots, 0]$ \Comment{Number of unsat constraints per point}
        \ForAll{triple of indices $1 \leq i < j < k \leq |P|$}
            \If{$\tau(i, j, k) \neq \sigma(p_i, p_j, p_k)$}
                \State $u \gets u + 1$
                \State $F[i], F[j], F[k] \gets F[i] + 1, F[j] + 1, F[k] + 1$
            \EndIf
        \EndFor
        \State \Return $u, F$
    \EndFunction
    \\
    \Function{LocalEval}{$P, \tau, i$}
        \State $F_i \gets [0, 0, \ldots, 0]$ \Comment{Number of unsat constraints involving $p_i$ per point}
        \ForAll{triple of indices $1 \leq a < b < c \leq |P|$ with $i \in \{a, b, c\}$}
            \If{$\tau(a, b, c) \neq \sigma(p_a, p_b, p_c)$}
                \State $F_i[a], F_i[b], F_i[c] \gets F_i[a]+1, F_i[b]+1, F_i[c]+1$
            \EndIf
        \EndFor
        \State \Return $F_i$
    \EndFunction
    \\
    \Function{WeightedSample}{$A, W$}
        \State randomly sample $i \in \{1, \ldots, |A|\}$ with probability proportional to $W[i] + 1$.
        \Comment{The $+1$ ensures that all elements can be chosen.}
        \State \Return $A[i]$
    \EndFunction

\end{algorithmic}
\end{algorithm}

\begin{algorithm}
    \newcommand{\itsSinceCheckpoint}{\textsf{itsSinceCheck}}
    \newcommand{\moveIdx}{i}
    \caption{\localizer~Thread}
    \label{alg:localizer_thread}
    \begin{algorithmic}[1]
    \Require An orientation assignment $\tau : \binom{n}{3} \to \{-1, 1\}$.

    \State $P \sim \mathcal{U}([0, 1]^2)^n$ \Comment{Start with $n$ uniformly random points in $[0, 1]^2$.}
    \State $u, F \gets \textsc{Eval}(P, \tau)$ \Comment{$u$ is the total number of unsat constraints, $F$ is an array with the number of unsat constraints per point}
    \\
    \State $\itsSinceCheckpoint \gets 0$ \Comment{Number of iterations since last improvement}
    \While{$u > 0$}
        \LComment{ Choose the index of a point to move proportionally to unsat constraints}
        \State $\moveIdx \gets \textsc{WeightedSample}(\{1, \ldots, n\}, F)$ 
        \State $p \gets P[\moveIdx]$
        \State $F_i \gets \textsc{LocalEval}(P, \tau, \moveIdx)$ \Comment{Evaluate w.r.t. point $p$}\\
        \LComment{Move chosen point with exponentially decreasing radius}
        \For{$s \in \{0, \ldots, \textsf{ptMovements}\}$}
            \State $r \gets \max\{\textsf{minRadius}, \textsf{maxRadius}/2^s \}$
            \State $p' \sim \mathcal{U}(B(p, r))$ \Comment{sample from radius-$r$ ball around $p$} 
            \State $P' \gets P \setminus \{p\} \cup \{p'\}$
            \State $F_i' \gets \textsc{localEval}(P', \tau, \moveIdx)$
            \If{$F_i'[i] \leq F_i[i]$}
                \State $P \gets P'$ \Comment{Accept new point even if no strict improvement}
                \ForAll{$j \in \{1, \ldots, n\}$} \Comment{Update unsat per point}
                        \State $F[j] \gets F[j] + (F_i'[j] - F_i[j])$ 
                \EndFor
                \If{$F_i'[i] < F_i[i]$} \Comment{Strict improvement case}
                    \State $u \gets u - (u_i - u_i')$ \Comment{Update total number of unsat constraints}
                    
                    \State $\textsc{BroadcastToLeaderboard}(P, u)$
                    \State $\itsSinceCheckpoint \gets 0$
                \EndIf
            \EndIf
            \If{$u = 0$}
                \State \textbf{break}
            \EndIf
        \EndFor
        \\
        \State $\itsSinceCheckpoint \gets \itsSinceCheckpoint + 1$
        \LComment{If no improvement in a while, restart from a good solution.}
        \If{$\itsSinceCheckpoint > \textsf{restartThreshold}$}
            \State $\textsf{newStart} \gets \textsc{WeightedSample}(\textsf{leaderboardSols}, \textsf{leaderboardScores})$ 
            \LComment{Perturb solution to increase diversity}
            \ForAll{$i \in \{1, \ldots, n\}$}
                \State $P[i] \sim \mathcal{U}(B(\textsf{newStart}[i], \textsf{resetRadius}))$
            \EndFor
            \State $u, F \gets \textsc{Eval}(P, \tau)$
        \EndIf 

    \EndWhile
    \State \Return $P$
\end{algorithmic}
\end{algorithm}

\end{document}